\documentclass[11pt]{article}
\usepackage{amsmath,amssymb}
\usepackage{framed}

\usepackage{graphicx}

\usepackage{amsmath,latexsym}

\begin{document}

\setcounter{page}{1}

\pagestyle{plain}

\begin{center}
\Large{\bf Mimetic DBI Inflation in Confrontation with Planck2018 data }\\
\small \vspace{1cm} { Kourosh
Nozari}$^{a,b,}$\footnote{knozari@umz.ac.ir} \quad and \quad {
Narges Rashidi}$^{a,}$\footnote{n.rashidi@umz.ac.ir} \\
\vspace{0.5cm} $^{a}$Department of Physics, Faculty of Basic
Sciences,
University of Mazandaran,\\
P. O. Box 47416-95447, Babolsar, IRAN\\
$^{b}$ Research Institute for Astronomy and Astrophysics of Maragha (RIAAM),\\
P. O. Box 55134-441, Maragha, Iran\\
\end{center}

\begin{abstract}
We study mimetic gravity in the presence of a DBI-like term which is
a non-canonical setup of the scalar field's derivatives. We consider
two general cases with varying and constant sound speeds and
construct the potentials for both the DBI and Mimetic DBI models. By
considering the power-law scale factor as $a=a_{0}\,t^{n}$, we seek
for the observational viability of these models. We show that, the
Mimetic DBI model in some ranges of the parameters space is free of
ghost and gradient instabilities. By studying $r-n_{s}$ and
$\alpha_{s}-n_{s}$ behavior in confrontation with Planck2018 data,
we find some constraints on the model's parameters. We show that
for the case with varying sound speed, although power-law DBI inflation is not
consistent with Planck2018 TT, TE, EE+low E+lensing data, but the
Mimetic DBI inflation is consistent with Planck2018
TT, TE, EE+low E+lensing data at 95$\%$ CL, in some ranges of the
model's parameters space as $40\leq n \leq 55$ where the
model is instabilities-free in these ranges of parameters too.
For the constant sound speed, by adopting some sample values of $c_{s}$, we study both DBI and
Mimetic DBI model numerically and find $n\sim 10^{2}$ for DBI model
and $n\sim 10$ for Mimetic DBI model. We also compare the
results with Planck2018 TT, TE, EE+low E+lensing+BK14+BAO data and see
that the DBI and Mimetic DBI model with varying sound speed are
ruled out with these joint data. However, these models with constant
sound speed are consistent with Planck2018 TT, TE, EE+low E+lensing+BK14+BAO data with $n\sim 10^{2}$ for DBI
model and $n\sim 10$ for Mimetic DBI model. In this case,
we find some tighter constraints on the corresponding sound speed.
\end{abstract}
\newpage

\section{Introduction}

In 2013, Chamseddine and Mukhanov have proposed a new approach to
General Relativity, respecting the conformal symmetry as an internal
degree of freedom~\cite{Cham13}. In their proposal, the authors have
introduced a model in which the physical metric is written in terms
of an auxiliary metric and a scalar field as follows
\begin{eqnarray}\label{eq1}
g_{\mu\nu}= -\tilde{g}^{\alpha\beta}\,\phi_{,\alpha}\,\phi_{,\beta}
\,\tilde{g}_{\mu\nu}\,.
\end{eqnarray}
The scalar field $\phi$ is a free and non-dynamical field which encodes the conformal mode of the
gravity. Considering the definition given in
equation (\ref{eq1}), we see that the physical metric ($g_{\mu\nu}$)
remains invariant with respect to the Weyl transformation of the
auxiliary metric ($\tilde{g}_{\mu\nu}$). Note that, the definition
(\ref{eq1}) leads to the following constraint on the scalar
field~\cite{Cham13}
\begin{equation}\label{eq2}
g^{\mu\nu}\phi_{,\mu}\phi_{,\nu}=-1\,,
\end{equation}
In Ref.~\cite{Cham13}, the action consists of the Einstein-Hilbert term
and the contribution of the matter coupled to $g_{\mu\nu}$. The
Einstein's field equations in this model consist of an extra
term which corresponds to the mimetic field. This extra term is
considered as a source of the dark matter. Actually, in the energy
density, there is a term corresponding to $a^{-3}$ which ``mimics" the dark
matter. In Ref.~\cite{Gol13}, by using the Lagrange multipliers in the
action, the author has introduced another mathematical approach to
explore the mimetic matter scenario (see also~\cite{Ham13}). The ghost free
models of the mimetic dark matter theory have been discussed
in~\cite{Bar13}. The lagrange multiplier approach has been followed
in~\cite{Cham14} where the authors have considered the Lagrange
multiplier in the action and also a potential for the mimetic field.
The authors in this paper have shown that by adopting the appropriate
potentials, it is possible to consider the mimetic field as
quintessence, phantom or inflaton fields. Other extensions of the
mimetic model, such as the non-minimal mimetic
model~\cite{Myr15,Hos18}, braneworld mimetic model~\cite{Sad17} and
mimetic gravity in the spirit of $f(R)$
theories~\cite{Noj14,Ast16,Noj16}, mimetic $f(G)$ gravity~\cite{Ast15},
mimetic Horndeski gravity~\cite{Co16,Arr15}, mimetic Galileon
gravity~\cite{Hag14}, unimodular mimetic $f(R)$ gravity~\cite{Odi16} and
mimetic Born-Infeld theory~\cite{Bou17,Che17} have been studied.

In the ordinary mimetic model, where even a potential is included,
the sound speed of the perturbations is vanishing and the
perturbations behave like a dust~\cite{Cham14}. In this regard, adding
higher derivative terms such as $\gamma\square \phi$ (where $\gamma$ is
a constant) leads to a non-zero sound speed of the
perturbations~\cite{Cham14}. By appropriate choosing of $\gamma$,
the mimetic model with $\gamma\square \phi$ term can be ghost free,
however, it still suffers from the gradient instability~\cite{Ij16}.
The authors of~\cite{Zhe17} have shown that by considering the
direct couplings of the higher derivatives of the mimetic field to
the curvature of the space-time, it is possible to overcome such
instabilities in some ranges of the parameters space. For more
papers on the (in)stability issue
see~\cite{Cap14,Mir14,Mal14,Ram15,Lan15,Ram16,Arr16,Ac16,Hir17,Cai17,Cai17,Tak17,Yos17}.

In this paper, instead of including the higher order derivatives of
the scalar field, we consider a non-canonical setup of the scalar
field's derivatives. DBI~\cite{Sil04,Che07,Ali04,Che05} and
tachyon~\cite{Se99,Se02a,Se02b,Noz13,Noz14,Noz18} fields are
examples of the non-canonical scalar fields. We add a DBI like term,
$f^{-1}(\phi)\,\sqrt{1-f\dot{\phi}^{2}}$, in the action of the
simple mimetic model with potential. Here, the field $\phi$ in DBI
term is the mimetic scalar field. In the string based DBI
(Dirac-Born-Infeld) model, the scalar field is characterized by the
radial position of a $D3$ brane~\cite{Sil04,Ali04}. This brane moves
in a ``throat'' region of a warped compactified space while there is
a speed limit on its motion~\cite{Sil04}. This scalar field can be
considered as inflaton with non-canonical kinetic term. In the DBI
inflation model the sound speed of the primordial perturbation is
different from unity (considering $c$, the speed of light, to be
unity) which can lead to large
non-Gaussianity~\cite{Ali04,Che05,Li08}. For more work on the DBI
inflation see~\cite{Che05b,Spa07,
Noz13a,Li14,Naz16,Qiu16,Kum16,Cho15,Ras18,Che18,ama18,Soul18}.
However, in this paper we assume the field in the DBI term to be the
mimetic scalar field. We are going to explore the cosmological
dynamics of the mimetic model in the presence of the DBI-like term.
We wonder, by adding this term, if the the mimetic model would be
stable- meaning that, it would be free of the gradient and ghost
instabilities or not. We also seek for the observational viability
of the model. To this end, we study the Mimetic DBI (from now on,
MDBI) model with power-law scale
factor~\cite{Luc85,Hal87,Lid89,Spa07b,Un13,Spa07}. In this regard we
consider a mimetic model with DBI-like term and potential, and also
use the lagrange multiplier in the action. We follow
Refs.~\cite{Bam14,Odi15} to find the appropriate potential and
lagrange multiplier corresponding to the power-law scale factor, and
then check the observational viability of the model.

This paper is organized as follows: In section 2, we consider the
case with varying sound speed. In subsection 2.1, we consider an
ordinary DBI model and obtain the main background equations of the
model. By introducing the slow-roll parameters and the sound speed
of the model, we obtain the main perturbation parameters like as
the tensor-to-scalar ratio, scalar spectral index and its running. We
also construct the potential of the model in terms of the Hubble
parameter and its derivatives. In subsection 2.2 we consider a
power-law scale factor and recast the slow-roll parameters (and
therefore the perturbation parameters) of the DBI model in terms of
the new parameters of the introduced scale factor. Given that, for the DBI
model we obtain two forms of potential and we analyze the perturbations
parameters for both cases. As we shall see, the DBI model that we
study here, is not observationally viable. But this not the end of the road. In section 2.3, we
introduce the Mimetic DBI inflation. In this section, by using the
Lagrange multiplier, we enter the mimetic constraint. In this
section also, we obtain the background equations of the model. By
using the Friedmann equations, we obtain the Lagrange multiplier and
the potential in terms of the Hubble parameter and its derivatives.
We also find the slow-roll parameters in our MDBI model which
differs from the ones in usual DBI model. In section 2.4, we study the
power-law MDBI model. We show that, by adding a DBI-like term in the
action of the mimetic model with potential term, and by considering
a power-law scale factor, the mimetic model would be free of gradient
instability in some ranges of the model's parameters space. We show
that the model is free of ghost instability in some ranges of the
model's parameters space which mach with the ranges leading to
the gradient instability free MDBI model. We also explore the
perturbation parameters numerically. The interesting point is that,
$r-n_{s}$ and $\alpha_{s}-n_{s}$ are consistent with observational
data in ranges of the model's parameters space where the MDBI model
is stable. Although these ranges lead to small non-Gaussianity, but
it is consistent with observational data. We use the constraints
obtained from $r-n_{s}$ and $\alpha_{s}-n_{s}$ to find the viable values
of the non-Gaussianity. In section 3 we use the constraint on the
nonlinearity parameter (in the non-Gaussianity) and constraint on
the sound speed to explore other features of the model. In this regard, we consider
constant sound speed and adopt some observationally viable values of
this quantity. By using the adopted values of the sound speed, we study
$r-n_{s}$ and $\alpha-n_{s}$ for DBI model in subsection 3.1 and
MDBI model in subsection 3.2 and find some constraints on the
model's parameters. In section 4, we consider the B-mode
polarization and use the BICEP2/Keck Array 2014 and Planck2018 joint
data to explore the models. As another important result, the DBI and MDBI models with power-law
scale factor and varying sound speed are ruled out by the
BICEP2/Keck Array 2014 and Planck2018 joint data. Therefore, in
section 4 we only perform numerical analysis on these models with
constant sound speed and find some constraints on the model. In
section 5, we present the summary and conclusion.

\section{Varying Sound speed}

In this section, we consider the case with varying sound speed
where the sound speed is expressed in terms of the model's
parameters. In this case, by using the observational viability of
$n_{s}$, $r$ and $\alpha_{s}$, we obtain some constraints on $n$ (remember that $a=a_{0}\,t^{n}$).
Then, by using the constraints on $n$, we obtain the observational
viable values of $c_{s}$ and the prediction of the model for
non-Gaussian features.

\subsection{DBI Model}

We consider the following action for the DBI model
\begin{eqnarray}
\label{eq3} S=\int
d^{4}x\sqrt{-g}\Bigg[\frac{1}{2\kappa^{2}}R-f^{-1}(\phi)\,\sqrt{1-2f(\phi)X}-V(\phi)
\Bigg],
\end{eqnarray}
where, $R$ is the Ricci scalar, $V(\phi)$ is the potential of the
field, $f^{-1}(\phi)$ is the inverse brane tension, related to the
geometry of the throat of the compact manifold passed by the
D3-brane and $X=-\frac{1}{2}\partial_{\nu}\phi\,\partial^{\nu}\phi$.

Variation of the action (\ref{eq3}) with respect to the metric leads to
the following Einstein's field equations
\begin{eqnarray}
\label{eq4}
G_{\mu\nu}=\kappa^{2}\Bigg[-g_{\mu\nu}f^{-1}\sqrt{1+f\,g^{\mu\nu}\,\partial_{\mu}\phi\,\partial_{\nu}\phi}-g_{\mu\nu}V
+\partial_{\mu}\phi\,\partial_{\nu}\phi
\Big(1+f\,g^{\mu\nu}\,\partial_{\mu}\phi\,\partial_{\nu}\phi\Big)^{-\frac{1}{2}}\Bigg]\,,
\end{eqnarray}
which by considering the flat FRW metric
\begin{equation}
\label{eq5} ds^{2}=-dt^{2}+a^{2}(t)\delta_{ij}dx^{i}dx^{j}\,,
\end{equation}
lead to the following Friedmann equations
\begin{eqnarray}
\label{eq6}
3H^{2}=\kappa^{2}\Bigg[\frac{f^{-1}}{\sqrt{1-f\dot{\phi}^{2}}}+V\Bigg]\,,
\end{eqnarray}

\begin{eqnarray}
\label{eq7}
2\dot{H}+3H^{2}=\kappa^{2}\Bigg[f^{-1}\,\sqrt{1-f\dot{\phi}^{2}}+V\Bigg]\,.
\end{eqnarray}
Variation of the action (\ref{eq3}) with respect to $\phi$ gives
the equation of motion as~\cite{Ras18}
\begin{equation}
\label{eq8}\frac{\ddot{\phi}}{(1-f\dot{\phi}^{2})^{\frac{3}{2}}}+\frac{3H\dot{\phi}}{(1-f\dot{\phi}^{2})^{\frac{1}{2}}}
+V'=-\frac{f'}{f^{2}}
\Bigg[\frac{3f\dot{\phi}^{2}-2}{2(1-f\dot{\phi}^{2})^{\frac{1}{2}}}\Bigg]\,.
\end{equation}

In this paper we are going to study the cosmic inflationary phase in this setup. To have
the inflation phase, the slow-roll parameters should be much smaller
than unity. These parameters are defined as
\begin{equation}
\label{eq9}\epsilon\equiv-\frac{\dot{H}}{H^{2}}\,,\quad
\eta=-\frac{1}{H}\frac{\ddot{H}}{\dot{H}}\,,\quad
s=\frac{1}{H}\frac{d \ln c_{s}}{dt}\,,
\end{equation}
where $c_{s}$ is the sound speed of the perturbations. To have inflation phase, the constraints
$f\dot{\phi}^{2}\ll 1$ and $\ddot{\phi}\ll 3H\dot{\phi}$ should be
satisfied.

In this model, the square of sound speed which is defined as
$c_{s}^{2}=\frac{P_{,X}}{\rho_{,X}}$ (where the subscript ``$,X$" shows
derivative with respect to $X$), is given by
\begin{eqnarray}
\label{eq10} c_{s}^{2}=1-f\dot{\phi}^{2}\,.
\end{eqnarray}
The sound speed should satisfy two constraints as
follows~\cite{Eli07,Qui17}. To avoid the appearance of gradient
instabilities, the square of the sound speed of the perturbation
should be positive, that is, $c_{s}^{2}>0$. Also, by considering
the causality requirement, the maximum value of the sound speed
should be equal to the value of the local speed of light- that
means $c_{s}\leq c$, where $c$ is the local speed of light. By
adopting $c\equiv 1$, these constraints imply that $0<c_{s}^{2}\leq
c^2$.

In this paper, we are going to seek for the observational viability
of the model by comparing the results of our model with the
Planck2018 observational data~\cite{pl18a,pl18b} and constraining
the perturbation parameters. To constraint the perturbation
parameters, Planck collaboration has used this fact that the
two-point correlations of the CMB anisotropies are described by the
angular power spectra $C_{l}^{TT}$, $C_{l}^{TE}$, $C_{l}^{EE}$ and
$C_{l}^{BB}$ ($l$ is the multipole moment number), under the assumption of
statistical isotropy~\cite{kam97,zal30,sel97,Hu97,Hu98}. Via the
transfer functions $\Delta_{l,A}^{s}$ and $\Delta_{l,A}^{T}$ (whic
are generally computed by using the Boltzmann codes such as
CMBFAST~\cite{sel96} or CAMB~\cite{Lew00}), the Planck collaboration
has related the CMB angular power spectra to the scalar and tensor
primordial power spectra. If we consider $a,b=T,E,B$, the
contributions from scalar and tensor perturbations in the  CMB
angular power spectra are given by~\cite{pl15}
\begin{eqnarray}
\label{eq11} C_{l}^{ab,s}=\int_{0}^{\infty} \frac{dk}{k}
\Delta_{l,a}^{s} (k) \, \Delta_{l,b}^{s} (k)\, {\cal{A}}_{s} (k)\,,
\end{eqnarray}

\begin{eqnarray}
\label{eq12} C_{l}^{ab,T}=\int_{0}^{\infty} \frac{dk}{k}
\Delta_{l,a}^{T} (k) \, \Delta_{l,b}^{T} (k)\, {\cal{A}}_{T} (k)\,.
\end{eqnarray}
The linear transformations encoded in $\Delta_{l,B}^{i} (k)$ (with
$i=s,T$) are corresponding to the physics of the late time and the
primordial power spectra ${\cal{A}}_{i}(k)$ are identified by the
physics of the primordial universe~\cite{pl15}. To explore the
transfer functions, and seek for the CMB anisotropies, the Planck
team has adopted the $\Lambda$CDM model as the one governing on the
late time background dynamics of the universe. Also, to compare the
perturbations parameters with data, they expand the scalar and tensor
power spectra as~\cite{pl18b,pl15} \footnote{Actually, the Planck
collaboration has adopted three procedures to compare the
inflationary parameters with data. One procedure is corresponding to
the parameterization of the primordial spectra as equations
(\ref{eq13}) and (\ref{eq14}). Another one is corresponding to the
dependence of the slow-roll power spectra on the Hubble
flow-functions. The third procedure is numerical and uses the
numerical codes like the CLASS~\cite{Les11,Bla11} and
ModeCode~\cite{Ada01,Pei03}. For more details see~\cite{pl15}.}
\begin{eqnarray}
\label{eq13} {\cal{A}}_{s}
(k)=A_{s}\left(\frac{k}{k_{*}}\right)^{n_{s}-1+\frac{1}{2}\frac{dn_{s}}{d\ln
k}\ln\big(\frac{k}{k_{*}}\big)+\frac{1}{6}\frac{d^{2}n_{s}}{d\ln
k^{2}}\ln\big(\frac{k}{k_{*}}\big)^{2}+...}\,,
\end{eqnarray}
\begin{eqnarray}
\label{eq14} {\cal{A}}_{T}
(k)=A_{T}\left(\frac{k}{k_{*}}\right)^{n_{T}+\frac{1}{2}\frac{dn_{T}}{d\ln
k}\ln\big(\frac{k}{k_{*}}\big)+...}\,,
\end{eqnarray}
which are model independent and where, $A_{i}$ is the amplitude of
the scalar ($i=s$) or tensor ($i=T$) perturbations. Also,
$\frac{dn_{i}}{d\ln k}$ is the the running of the scalar ($i=s$) or
tensor ($i=T$) spectral index and $\frac{d^{2}n_{s}}{d\ln k^{2}}$ is
the running of the running of the scalar spectral index. Using the
power spectra, it is possible to find the ratio of the tensor to
scalar amplitudes of the perturbations as
\begin{eqnarray}
\label{eq15} r=\frac{{\cal{A}}_{T}(k_{*})}{{\cal{A}}_{s}(k_{*})}\,,
\end{eqnarray}
which is an important perturbation parameter. In this paper, we
consider the constraints on the perturbation parameters obtained by
Planck2018 team, based on the $\Lambda$CDM$+r+\frac{dn_{s}}{d\ln k}$
model which supports quasi-de Sitter expansion during inflation
epoch. In the $\Lambda$CDM$+r+\frac{dn_{s}}{d\ln k}$ model, the
amplitude and the scale dependence of the tensor perturbations, the
amplitude and the scale dependence of the scalar perturbations and
the scale dependence of the scalar spectral index are taken into
account. From the Planck2018 TT, TE, EE+ low EB+ lensing data, the
constraint on the scalar spectral index in
$\Lambda$CDM$+r+\frac{dn_{s}}{d\ln k}$ model is as $n_{s}=0.9647\pm
0.0044$, the constraint on the tensor-to-scalar ratio is as $r<0.16$
and the constraint on the running of the scalar spectral index is as
$\frac{dn_{s}}{d\ln k}=-0.0085\pm 0.0073$, which are calculated at
pivot scale $k_{*}=0.002 Mpc^{-1}$~\cite{pl18b}. Note that,
Planck2018 team has obtained some constraints based on
$\Lambda$CDM$+r$ model too ($n_{s}=0.9659\pm 0.0041$ and $r<0.10$
from Planck2018 TT, TE, EE+ low EB+ lensing), however, here we focus on the
results of $\Lambda$CDM$+r+\frac{dn_{s}}{d\ln k}$
model~\cite{pl18b}.

Now, we should obtain the perturbation parameters in the DBI model.
Following the Planck collaboration assumption, we assume the
late-time background dynamics is governed by $\Lambda$CDM model and
the running of the spectral index is taken into account. The scale
dependence of the scalar spectral index is identified by
\begin{eqnarray}
\label{eq16} n_{s}-1=\frac{d \ln {\cal{A}}_{s}}{d\ln
k}\Bigg|_{c_{s}k=aH}
\end{eqnarray}
In this equation, the subscript $c_{s}k=aH$ means that the scalar
spectral index is calculated at the time of sound horizon exit of
the physical scales. Note that, although we consider the running of
the scalar spectral index, but since the inflationary parameters are
calculated at pivot scale $k=k_{*}$, the running term doesn't appear
in definition (\ref{eq16}). The amplitude of the scalar spectral
index, which Planck2018 TT, TE, EE+ low EB+ lensing data gives its
value as ${\cal{A}}_{s}\simeq 2.0989\times10^{-9}$~\cite{pl18b}, is
given by (see~\cite{Ras18} for instance)
\begin{equation}
\label{eq17}{\cal{A}}_{s}=\frac{H^{2}}{8\pi^{2}{\cal{W}}_{s}c_{s}^{3}}\,,
\end{equation}
where
\begin{equation}
\label{eq18} {\cal{W}}_{s}=\frac { \dot{\phi}^{2}}{2{H}^{2} \left(
1-f\, \dot{\phi}^{2} \right) ^{3/2}}\,.
\end{equation}
To avoid the ghost instability, ${\cal{W}}_{s}$ should be positive.

We can write the scalar spectral index, in terms of the slow-roll
parameters, as follows
\begin{eqnarray}
\label{eq19} n_{s}=1-6\epsilon+2\eta-s\,.
\end{eqnarray}
As we have mentioned earlier, the Planck2018 TT, TE, EE+low EB+lensing
data gives the value of the scalar spectral index as
$n_{s}=0.9647\pm 0.0044$ (in the $\Lambda$CDM$+r+\frac{dn_{s}}{d\ln
k}$ model)~\cite{pl18b}. Another important parameter is the running
of the scalar spectral index which is given by
\begin{eqnarray}
\label{eq20}
\alpha_{s}=8\epsilon\,\eta-12\epsilon^{2}-2\zeta+2\eta^{2}-\frac{1}{H}\dot{s}\,,
\end{eqnarray}
where
\begin{eqnarray}
\label{eq21} \zeta=\frac{\dddot{H}}{H^{2}\dot{H}}\,.
\end{eqnarray}
The value of the running of the scalar spectral index, from
Planck2018 TT, TE, EE + low EB+ lensing data is as $\alpha_{s}=-0.0085\pm
0.0073$~\cite{pl18b}.

From equation (\ref{eq14}) we see that the tensor spectral index is
defined as follows
\begin{eqnarray}
\label{eq22} n_{T}=\frac{d \ln {\cal{A}}_{T}}{d\ln
k}\Bigg|_{k=aH}\,,
\end{eqnarray}
where, the subscript $k=aH$ means that the tensor spectral index is
calculated at Hubble horizon crossing of the physical scale. In
definition (\ref{eq22}), ${\cal{A}}_{T}$, the amplitude of the
tensor perturbations, is given by
\begin{eqnarray}
\label{eq23} {\cal{A}}_{T}=\frac{2\kappa^{2}H^{2}}{\pi^{2}}\,.
\end{eqnarray}
Now, we have the tensor spectral index as
\begin{eqnarray}
\label{eq24} n_{T}=-2\epsilon\,.
\end{eqnarray}
Considering that there is no detection of a non-zero tensor
amplitude by current data~\cite{pl18b}, when $r$ is very small
(actually enough close to zero), any values of the tensor
perturbations is viable in essence~\cite{pl18b}. By relaxing the inflationary
consistency relation and considering Planck2018 TT, TE, EE+ low E
+lensing+ BK14+ LIGO $\&$ Virgo2016 data the Planck team have obtained
the constraint on the tensor part as $-0.62<n_{T}<0.53$. However,
there is no exact value of $n_{t}$. In Ref.~\cite{Sim07}, one can see
the different delensing techniques and their ability to constraint
the tensor spectral index.

By using equations (\ref{eq15}) and (\ref{eq24}) we obtain the
tensor-to-scalar ratio as follows
\begin{eqnarray}
\label{eq25} r=16\,c_{s}\,\epsilon\,,
\end{eqnarray}
or
\begin{eqnarray}
\label{eq26} r=-8\,c_{s}\,n_{T}\,.
\end{eqnarray}
Equation (\ref{eq26}) is named the consistency relation. Note that
in the simple single filed inflation with a canonical scalar field
we have $c_{s}^{2}=1$ and therefore $r=16\epsilon$. However, in the
non canonical DBI field the tensor-to-scalar ratio is given by Eq.
(\ref{eq26}). In some extended models, there is some additional
terms in consistency relation. In other words, in those cases, the
consistency relation is
modified~\cite{Noz13,Noz14,Fel11a,Fel11b,Noz17}. As we said, the
constraint on the tensor-to-scalar ratio, from Planck2018 TT, TE,
EE+ lowEB + lensing data, is as $r<0.16$~\cite{pl18b}.

Now, following Refs.~\cite{Bam14,Odi15}, we use the Friedmann
equations to obtain the potential in terms of the Hubble parameter
and its derivatives. First of all, we adopt $f^{-1}(\phi)=V(\phi)$.
Then, we introduce a new scalar field $\varphi$ which is identified
by the number of e-folds $N$ and parameterizes the scalar field
$\phi$ as $\phi=\phi(\varphi)$. By using these points, equation
(\ref{eq7}) gives the following expression for the potential
\begin{eqnarray}
\label{eq27} V_{\pm}=-{\frac {{12H}^{2}(N)\,H'(N)}{{\kappa}^{2}
\left( \pm\sqrt {3}\sqrt {H(N)} \sqrt
{3\,H(N)+8\,H'(N)}+3\,H(N)-4\,H'(N) \right) }}\,,
\end{eqnarray}
where a prime shows a derivative of the parameter with respect to
$N$. We also find the following expressions for the slow-roll
parameters in terms of the Hubble parameter and its derivatives
\begin{eqnarray}
\label{eq28}\epsilon_{\pm}={\frac { \left( H^{3}H''
+H^{2}H'^{2}+\frac{4}{3} H^{2}H'H''+4HH'^{3}\pm \frac{1}{\sqrt{3}} e
H^{\frac{5}{2}}H'' \pm \frac{1}{\sqrt{3}} eH^{\frac{3}{2}}H'^{2}\mp
\frac{8}{3\sqrt{3}}e H'^{3}\sqrt {H } \right) ^{2}}{-\frac{4}{27} H
^{2} \left( \sqrt {3}e\sqrt {H }+3\,H -4 H' \right) ^{2} H'^{3}
\left( 3\,H +8\,H' \right) }}\,,
\end{eqnarray}
where, $e\equiv\sqrt {H+8\,H'}$ and $H\equiv H(N)$.

\begin{eqnarray}
\label{eq29}\eta_{\pm}=- \Bigg\{ \frac{1}{4V_{\pm}^{2}} \Bigg[
-\frac{{\kappa}^{2}}{2} \left( 1-{\frac {HH'' }{ H'^{2}}} \right)
V'_{\pm} -{\frac {{\kappa}^{2}H V''_{\pm} }{H' }} -2 \Bigg(
\frac{{\kappa}^{2}}{4} \left( 1-{\frac {H H'' }{ H'^{2}}} \right)
\frac{V'_{\pm}}{V_{\pm}^{2}}\nonumber\\-\frac{\kappa^{2}}{2}\frac{H}{H'}
\left( 2{\frac { V_{\pm}'^{2}}{ V_{\pm}^{3}}}-{ \frac {V''_{\pm} }{
V_{\pm}^{2}}} \right) \Bigg) V_{\pm}^{ 2} \Bigg]+\frac{1}{2}{\frac {
V_{\pm}'^{2}{\kappa}^{2}H }{ H'V_{\pm}^{2}}} \Bigg\} {\kappa}^{
-2}\,,
\end{eqnarray}

\begin{eqnarray}
\label{eq30}s=\frac{1}{2H} \Bigg\{ -2{\frac { V'_{\pm} H^{2}\,H' }{
V_{\pm}^{2}\,{\kappa}^{ 2}}}-2 \frac{H^{2}}{V_{\pm}} \Bigg[ - \left(
{ \frac {H''  }{H  }}-{\frac { H'^{2}}{ H^{2}}} \right) {\frac
{{\kappa}^{-2}H  }{\sqrt {-{\frac {2H' }{{\kappa}^{2}H
}}}}}\nonumber\\+ {\frac {H' }{\sqrt {-{\frac {{ \kappa}^{2}H  }{2H'
}}}} } \Bigg]{\frac {1}{\sqrt {-{\frac {{ \kappa}^{2}H }{2H'}}}}}
\Bigg\} \left( 1+2\,{\frac {H H' }{V_{\pm} {\kappa}^{2}}} \right)
^{-1}\,.
\end{eqnarray}
Now, the perturbation parameters $n_{s}$ and $r$ can be expressed in
terms of the Hubble parameter and its derivatives. By adopting some
functions for the Hubble parameter, corresponding to some inflation
model, we can analyze the observable parameters numerically.

Note that the tensor-to-scalar ratio is corresponding to the sound
speed. This parameter is related to the other important property in
the inflationary models, named ``non-Gaussianity''. This property
which is generated during the inflation era, can be used to test the
observational viability of the inflation models. The Gaussian
perturbations are characterized by two-point correlation. However,
the additional statistical information corresponding to the
non-Gaussian distribution can be obtained from three and higher
order correlations. The 3-point correlation in the interaction
picture is defined as~\cite{Mal03}
\begin{eqnarray}
\label{eq31}<\Phi(\textbf{k}_{1})\,\Phi(\textbf{k}_{2}\,\Phi(\textbf{k}_{3}))=(2\pi)^{7}\,\delta^{3}(\textbf{k}_{1}
+\textbf{k}_{2}+\textbf{k}_{3}){\cal{B}}_{\Phi}(k_{1},k_{2},k_{3})\,,
\end{eqnarray}
where the potential $\Phi$, which is equivalent to the Bardeen
gravitational potential, is defined as $\Phi=\frac{3}{5}\xi$ with
$\xi$ to be the co-moving curvature perturbation. As we see from
equation (\ref{eq31}), the bispectrum ${\cal{B}}_{\Phi}$ depends on
the three momenta $\textbf{k}_{1}$, $\textbf{k}_{2}$ and $\textbf{k}_{3}$. By considering the
translational and rotational invariance and depending on the amount
of momenta, we are faced with different shapes of the
non-Gaussianity with different corresponding amplitudes. In studying
the non-Gaussian feature of the perturbation using the three-point
correlation function, the so-called ``nonlinearity parameter''
measuring the amplitude of the non-Gaussianity of the primordial
perturbations, is an important parameter which is related to the
sound speed (see~\cite{Ali04,Che05}). The Planck collaboration has
obtained a constraint on the nonlinearity parameter in DBI model. To
obtain this constraint, the Planck team has used the following
bispectrum of the primordial perturbations~\cite{Ali04,Che05}
\begin{eqnarray}
\label{eq32}
{\cal{B}}_{\Phi}(k_{1},k_{2},k_{3})=\frac{6{\cal{A}}^{2}f_{NL}}{(k_{1}k_{2}k_{3})^{3}}\,\frac{-3}{7(k_{1}+k_{2}+k_{3})^{2}}
\Bigg[\sum_{i} k_{i}^{5}+\sum_{i\neq
j}\Big(2k_{i}^{4}k_{j}-3k_{i}^{3}k_{j}^{2}\Big)\nonumber\\+\sum_{i\neq
j \neq
l}\Big(k_{i}^{3}k_{j}k_{l}-4k_{i}^{2}k_{j}^{2}k_{l}\Big)\Bigg]\,,
\end{eqnarray}
where the power spectrum of the potential $\Phi$,
${\cal{P}}_{\Phi}(k)=\frac{{\cal{A}}}{k^{4-n_{s}}}$, has been
normalized to ${\cal{A}}^{2}$. By constraining the non-separable
shape given by the above equation, the Planck collaboration has
obtained the value of the nonlinearity parameter as $f_{NL}=15.6\pm
37.3$ from temperature and polarization data at 68\% CL~\cite{pl15}.
The nonlinearity parameter of the DBI model, given in the above
bispectrum, is related to the sound speed of the perturbations.\\

In the slow-roll approximation, the important terms in the amplitude
of the non-Gaussianity are as follows~\cite{Che07}
\begin{eqnarray}
\label{eq33a} f_{NL}=f_{NL}^{\sigma}+f_{NL}^{c}
\end{eqnarray}
with
\begin{eqnarray}
\label{eq33b}
f_{NL}^{\sigma}=-\frac{5}{81}\left(\frac{1}{c_{s}^{2}}-1-\frac{2\sigma}{\Sigma}\right)+\Big(3-2c_{1}\Big)\frac{l\sigma}{\Sigma}\,,
\end{eqnarray}
and
\begin{eqnarray}
\label{eq33c}
f_{NL}^{c}=\frac{35}{108}\Bigg(\frac{1}{c_{s}^{2}}-1\Bigg)\,,
\end{eqnarray}
where $c_{1}\approx 0.577$ is the Euler constant and
\begin{eqnarray}
\label{eq33d} \sigma=X^{2}P_{,XX}+\frac{2}{3}X^{3}P_{,XXX}\,,
\end{eqnarray}
\begin{eqnarray}
\label{eq33d} \Sigma=XP_{,X}+2X^{2}P_{,XX}\,,
\end{eqnarray}
\begin{eqnarray}
\label{eq33d} l=\frac{\dot{\sigma}}{\sigma H}\,.
\end{eqnarray}
In the above equations ${,X}$ means derivative with respect to $X$
and a dot shows the derivative with respect to the time. If we
consider the DBI model in which
\begin{eqnarray}
\label{eq33e} P(X,\phi)=-f^{-1}(\phi)\sqrt{1-2Xf(\phi)}-V(\phi)\,,
\end{eqnarray}
the leading order contribution in $f_{NL}^{\sigma}$
vanishes~\cite{Che07}. In this regard, only equation (\ref{eq33c})
contributes in the amplitude of the non-Gaussianity.\\

Therefore, we have~\cite{Che07,Ali04,Che05}
\begin{eqnarray}
\label{eq33}
f_{NL}=-\frac{35}{108}\Bigg(\frac{1}{c_{s}^{2}}-1\Bigg)\,.
\end{eqnarray}
So, by having the values of the nonlinearity parameter, it is
possible to get some constraints on the sound speed. Indeed, the
Planck team by using the constraint on $f_{NL}$ at 95\% CL and
equation (\ref{eq33}) has obtained constraints on the sound speed at
95\% CL. The constraints are as $c_{s}\geq 0.069$ from temperature
data only at 95\% CL, and $c_{s}\geq 0.087$ from temperature and
polarization data at 95\% CL.

\subsection{Power-Law Inflation in DBI Model}

Now, we study the power law inflation which is described by the following scale factor
\begin{eqnarray}
\label{eq34}a=a_{0}\,t^n\,,
\end{eqnarray}
leading to the following Hubble parameter
\begin{eqnarray}
\label{eq35}H(N)=n\,e^{-\frac{N}{n}}\,.
\end{eqnarray}
Actually, the scale factor (\ref{eq34}) leads to $H=\frac{n}{t}$,
which in terms of the e-fold's number is written as (\ref{eq35}). By
adopting the power law inflation, the slow-roll parameters take the
following form
\begin{eqnarray}
\label{eq36}\epsilon_{\pm}={\frac { \left( 3\,{n}^{\frac{3}{2}}\sqrt
{3\,n-8}\pm3\,\sqrt {3}{n}^{2}+4\, \sqrt {n}\sqrt
{3\,n-8}\mp8\,\sqrt {3}n \right) ^{2}}{{n}^{2} \left( \sqrt {n}\sqrt
{3\,n-8}\sqrt {3}\pm3\,n\pm4 \right) ^{2} \left( 3\,n-8 \right)
}}\,,
\end{eqnarray}

\begin{eqnarray}
\label{eq37} \eta_{\pm}=4\,{\frac {\mp9\,\sqrt
{3}{n}^{3}\pm12\,{n}^{2}\sqrt {3}+9\,{n}^{\frac{5}{2}} \sqrt
{3\,n-8}\pm32\,\sqrt {3}n+8\,\sqrt {n}\sqrt
{3\,n-8}}{{n}^{\frac{3}{2}} \left( \sqrt {n}\sqrt {3\,n-8}\sqrt
{3}\mp3\,n\mp4 \right) ^{2}\sqrt {3\, n-8}}}\,,
\end{eqnarray}

\begin{eqnarray}
\label{eq38} s=0\,.
\end{eqnarray}

By using the above relations, the scalar spectral index, its running
and tensor-to-scalar ratio can be expressed in terms of $N$ and $n$
and explored numerically. However, note that the tensor-to-scalar
ratio is corresponding to the sound speed (see equations
(\ref{eq25}) and (\ref{eq26})). On the other hand, as we have stated
earlier, to avoid the instabilities, the sound speed of the
perturbations should be as $0<c_{s}^{2}\leq 1$. DBI power-law inflation
for some values of $n$ is free of the gradient instabilities and
also satisfies the causality requirement for both cases with $V_{+}$
and $V_{-}$. The left panel of figure 1 shows the evolution of the
square of the sound speed versus $n$ for the case with $V_{+}$. For
all values of $n\geq 8$, the sound speed is positive and also
smaller than unity. The right panel of figure 1 shows evolution of
${\cal{W}}_{s}$ versus $n$. As figure shows, for all values of
$n\geq 8$ this model is free of ghost instabilities too.

\begin{figure}[]
\begin{center}
\includegraphics[scale=0.37]{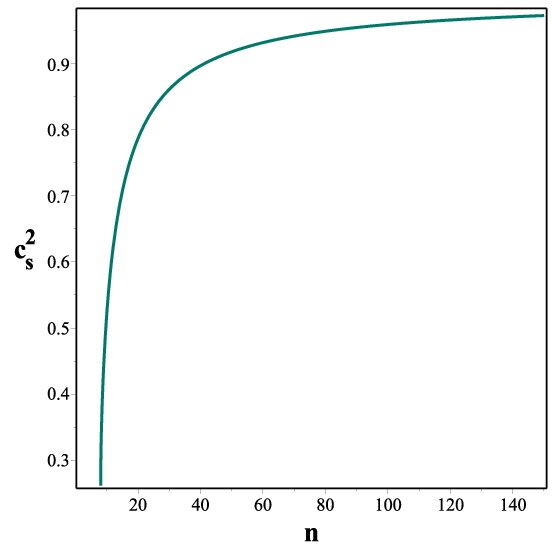}
\includegraphics[scale=0.37]{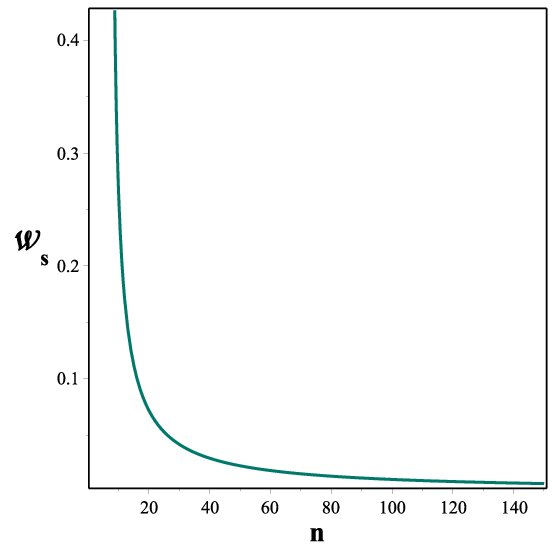}
\end{center}
\caption{\small {$c_{s}^{2}$ and ${\cal{W}}_{s}$} versus
$n$ in the DBI model with power-law scale factor as $a=a_{0}\,t^{n}$
and $V_{+}$. As figures show, for all values of $n$, the DBI model
is free of the gradient and ghost instabilities and also satisfies
the causality requirement.}
\label{fig1}
\end{figure}

Now, we find the constraints on the parameter $n$ by exploring the
observational viability of the tensor-to-scalar ratio, in
confrontation with the Planck2018 TT, TE, EE+lowEB+lensing data. By
using equations (\ref{eq10}), (\ref{eq25}) and (\ref{eq37}), for
$V_{+}$ we find
\begin{equation}
\label{eq39} r<0.16 \,\,\Big(\Lambda CDM +r+\frac{d n_{s}}{d \ln
k}\Big)\,\,\Longrightarrow \,\, n \geq 97.914\,.
\end{equation}
Now, from the constraint obtained on $n$ we can constraint the sound
speed of the model as
\begin{equation}
\label{eq40}  n\geq 97.914 \Longrightarrow \,\, c_{s}\geq
0.97914\,.
\end{equation}
By implying this constraint on equation (\ref{eq33}), we find
\begin{equation}
\label{eq41} c_{s} \geq 0.97914 \Longrightarrow \,\, -0.01395\leq
f_{NL}\leq 0\,.
\end{equation}
Therefore, in the power law DBI inflation the amplitude of the
non-gaussianity is small. However, the obtained range of $n$ doesn't
lead to the observationally viable values of the scalar spectral
index. Actually, considering the scalar spectral index in the
$\Lambda$CDM$+r+\frac{dn_{s}}{d \ln k}$ model, as $n_{s}=0.9647\pm
0.0044$ from Planck2018 TT, TE, EE+lowEB+lensing data, gives the
following constraint on $n$
\begin{equation}
\label{eq42} 50.378\leq n \leq 64.725\,.
\end{equation}
In fact, if we plot $r$-$n_{s}$ diagram in the background of
Planck2018 TT, TE, EE+lowE+lensing data, it is completely out of the
$68\%$ and $95\%$ confidence levels of the data set (see figure 2). In this regard, the power law inflation in DBI
model is not observationally viable. Also, in this case there is no
running of the scalar spectral index.

\begin{figure}[]
\begin{center}
\includegraphics[scale=0.3]{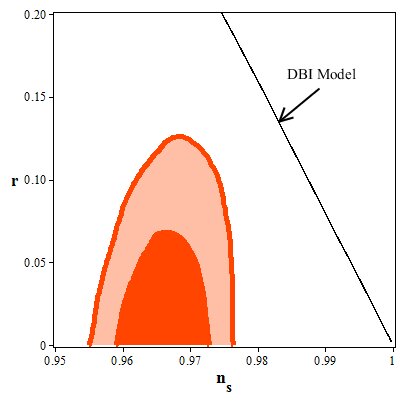}
\end{center}
\caption{\small {Tensor-to-scalar ratio versus the
scalar spectral index for DBI inflation with power law scale factor
as $a=a_{0}\,t^{n}$ and $V_{+}$ in the background of Planck2018
TT, TE, EE+lowE+lensing data. This model is not observationally
viable.}}
\label{fig2}
\end{figure}

We can also study the case corresponding to $V_{-}$. With this
potential, the behavior of the square of the sound speed is as the
left panel of figure 3. As figure shows, in this case also, the DBI
power law model with $V_{-}$, for $n\geq 8$ is free of the gradient
instabilities and also satisfies the causality requirement. The
right panel of figure 3 shows the ghost instability issue. As figure
shows, the DBI power law model with $V_{-}$ suffers from ghost
instability. By using equations (\ref{eq10}), (\ref{eq25}) and
(\ref{eq37}), for $V_{+}$ we find the following constraint on $n$
\begin{equation}
\label{eq43} r<0.16 \,\,\Big(\Lambda CDM +r+\frac{d n_{s}}{d \ln
k}\Big)\,\,\Longrightarrow \,\, n\geq 15.373\,.
\end{equation}
On the other hand, numerically exploring the tensor-to-scalar ratio
gives another constraint on $n$. By using equations (\ref{eq19}) and
(\ref{eq36})-(\ref{eq38}), and considering $n_{s}=0.9647\pm 0.0044$
from Planck2018 TT,TE,EE+lowEB+lensing data, we get
\begin{equation}
\label{eq44} 50.378\leq n \leq 64.725\,.
\end{equation}
This constraint limits the one obtained by exploring the
tensor-to-scalar ratio. Now, to obtain the constraint on the sound
speed of the perturbations in this model, we study $r-n_{s}$ plane
and find
\begin{equation}
\label{eq45}  45.1\leq n \leq 85.1 \Longrightarrow \,\,
0.04408 \leq c_{s} \leq  0.04651 \,.
\end{equation}
We note that, instead of using the constraint on $n$ and definition of
the sound speed, one can use relation (\ref{eq25}) and
observationally viable values of $r$ to obtain the observational
constraint on the sound speed~\cite{Noz18b}. This constraint on the
sound speed leads to the following constraint on the amplitude of
the non-Gaussianity
\begin{equation}
\label{eq46}  0.04408 \leq c_{s} \leq  0.04651
\Longrightarrow \,\, -9.63\times 10^{5} \leq f_{NL} \leq
-6.93\times 10^{4}\,,
\end{equation}
which is too large to be consistent with observational data.

\begin{figure}[]
\begin{center}
\includegraphics[scale=0.37]{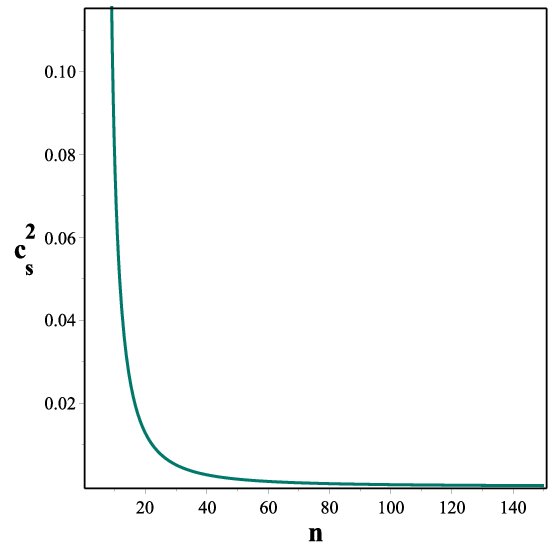}
\includegraphics[scale=0.37]{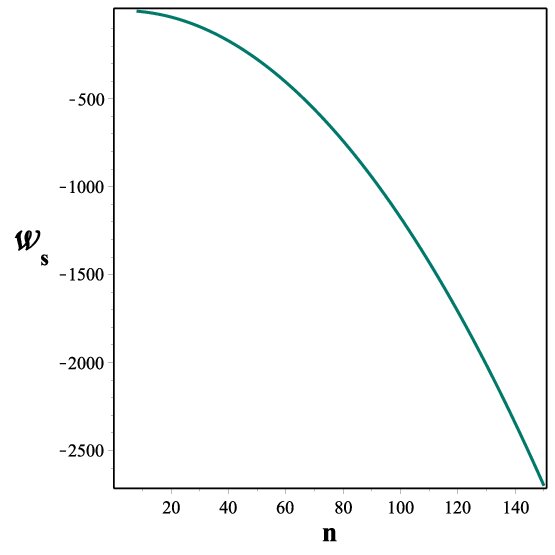}
\end{center}
\caption{\small {$c_{s}^{2}$ and ${\cal{W}}_{s}$} versus $n$ in the DBI model with power-law scale
factor $a=a_{0}\,t^{n}$ and $V_{-}$. As the left panel shows, for all
values of $n$, the DBI model is free of the gradient instabilities and also
satisfies the causality requirement. However, this model suffers
from the ghost instabilities as the right panel shows.}
\label{fig3}
\end{figure}

\begin{figure}[]
\begin{center}
\includegraphics[scale=0.3]{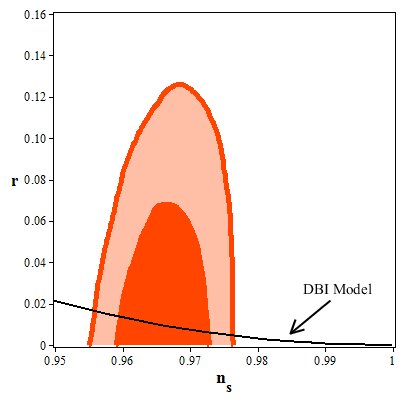}
\end{center}
\caption{\small {Tensor-to-scalar ratio versus the
scalar spectral index in the DBI model with power-law scale factor
as $a=a_{0}\,t^{n}$ and $V_{-}$ in the background of Planck2018
TT, TE, EE+lowE +lensing data. }}
\label{fig4}
\end{figure}

We conclude this section that the DBI model with action (\ref{eq3}) and a
power-law scale factor is not consistent with Planck2018
observational data. A question then arises: how can this tension be overcome?
In the next section, we are going to study the DBI
model in the\emph{ mimetic gravity} framework hoping to overcome this issue.
As we shall see, the Mimetic DBI (MDBI) model with power-law scale factor in some subspaces of the model parameter space is
consistent with Planck2018 TT, TE, EE+lowE+lensing data.

\subsection{Mimetic DBI Model}

We construct a DBI cosmological model in the spirit of mimetic gravity.
The action of this DBI mimetic gravity, by considering the Lagrange
multiplier and a potential term, is constructed as follows
\begin{eqnarray}
\label{eq47} S=\int
d^{4}x\sqrt{-g}\Bigg[\frac{1}{2\kappa^{2}}R-f^{-1}(\phi)\sqrt{1-2f(\phi)X}+\lambda(g^{\mu\nu}\partial_{\mu}\phi\,\partial_{\nu}\phi+1)-V(\phi)
\Bigg],
\end{eqnarray}
where, $\lambda$ is a Lagrange multiplier which lets us to enter the
mimetic constraint (\ref{eq2}) in the action. We note that one may
argue that this action cannot provide a propagating curvature
perturbation since if one exerts the mimetic constraint
$g^{\mu\nu}\partial_{\mu}\phi\partial_{\nu}\phi = -1$ in the DBI
term, this term effectively behaves like a potential term
($\sqrt{1-f(\phi)}$) and therefore the action (\ref{eq47}) is
equivalent to the original mimetic scenario. However, this is not
actually the case since in Lagrangian formalism one is not allowed
to impose the constraints on the action from the beginning. The
constraints should be imposed on equations of motion just after
derivation of the equations. With this point in mind, as we will
show action (\ref{eq47}) provides propagating curvature perturbation
($c_{s}\neq 0$). We note also that in the original mimetic scenario,
to have propagating curvature perturbation (with nonzero sound
speed) the authors added the higher derivative term $(\square
\phi)^{2}$ (see Ref.~\cite{Cham14} for details). Here we are able to
provide such a propagating modes in a mimetic DBI scenario without
additional $(\square \phi)^{2}$ term, provided that the mimetic
constraint to be imposed after derivation of the field equations.

Varying the action (\ref{eq47}) with respect to $\lambda$ recovers the
constraint equation (\ref{eq2}). Variation of the action with
respect to the metric leads to the following Einstein's field
equations
\begin{eqnarray}
\label{eq48}
G_{\mu\nu}=\kappa^{2}\Bigg[-g_{\mu\nu}f^{-1}\sqrt{1+f\,g^{\mu\nu}\,\partial_{\mu}\phi\,\partial_{\nu}\phi}-g_{\mu\nu}V
+g_{\mu\nu}\,\lambda\Big(g^{\mu\nu}\,\partial_{\mu}\phi\,\partial_{\nu}\phi+1\Big)\nonumber\\+\partial_{\mu}\phi\,\partial_{\nu}\phi
\Big(1+f\,g^{\mu\nu}\,\partial_{\mu}\phi\,\partial_{\nu}\phi\Big)^{-\frac{1}{2}}-2\lambda\,\partial_{\mu}\phi\,\partial_{\nu}\phi\Bigg]\,.
\end{eqnarray}
With the flat FRW metric (\ref{eq5}) the Einstein's field equations
(\ref{eq48}) lead to the following Friedmann equations
\begin{eqnarray}
\label{eq49}
3H^{2}=\kappa^{2}\Bigg[\frac{f^{-1}}{\sqrt{1-f\dot{\phi}^{2}}}+V-\lambda\Big(1+\dot{\phi}^{2}\Big)\Bigg]\,,
\end{eqnarray}

\begin{eqnarray}
\label{eq50}
2\dot{H}+3H^{2}=\kappa^{2}\Bigg[f^{-1}\,\sqrt{1-f\dot{\phi}^{2}}+V+\lambda\Big(\dot{\phi}^{2}-1\Big)\Bigg]\,.
\end{eqnarray}
Variation of the action (\ref{eq47}) with respect to $\phi$
gives the following equation of motion
\begin{equation}
\label{eq51}\frac{\ddot{\phi}}{(1-f\dot{\phi}^{2})^{\frac{3}{2}}}+\frac{3H\dot{\phi}}{(1-f\dot{\phi}^{2})^{\frac{1}{2}}}
-2\lambda\Big(\ddot{\phi}+3H\dot{\phi}\Big)+V'-\lambda'(1-\dot{\phi}^{2})=-\frac{f'}{f^{2}}
\Bigg[\frac{3f\dot{\phi}^{2}-2}{2(1-f\dot{\phi}^{2})^{\frac{1}{2}}}\Bigg]\,.
\end{equation}
The slow-roll parameters in the MDBI model are given by equations
(\ref{eq9}), however, the Hubble parameter and sound speed take new forms. The Hubble parameter is given by the Friedmann equation
(\ref{eq49}) and the square of sound speed is defined as
\begin{eqnarray}
\label{eq52}
c_{s}^{2}=\frac{\left(1-f\dot{\phi}^{2}\right)^{-\frac{1}{2}}-2\lambda}{\left(1-f\dot{\phi}^{2}\right)^{-\frac{3}{2}}-2\lambda}\,.
\end{eqnarray}

In the MDBI model, the tensor-to-scalar ratio and the scalar
spectral index are defined by equations (\ref{eq15}) and
(\ref{eq16}), where the amplitudes of the perturbations are given by
(\ref{eq17}) and (\ref{eq23}).\\

In fact, now the amplitude of the scalar perturbation is given by
\begin{equation}
\label{eq53a}{\cal{A}}_{s}=\frac{H^{2}}{8\pi^{2}{\cal{W}}_{s}c_{s}^{3}}=\frac{H^{2}}{8\pi^{2}({\cal{W}}_{s}^{(DBI)}+\lambda
W)c_{s}^{3}}\,,
\end{equation}
where we have used ${\cal{W}}_{s}={\cal{W}}_{s}^{(DBI)}+\lambda W$,
with
\begin{equation}
\label{eq53b} W=\frac {\left( f\,\dot{\phi}^{2}\,\sqrt
{1-f\dot{\phi}^{2}}-\,\sqrt {1-f\dot{\phi}^{2}} \right)
\dot{\phi}^{2}}{2{H}^{2} \left( 1-f\dot{\phi}^{2} \right) ^{3/2}}\,.
\end{equation}
and ${\cal{W}}_{s}^{(DBI)}$ is given by the equation (\ref{eq18}).\\

This means that in the MDBI model, the parameter ${\cal{W}}_{s}$ is
given by
\begin{equation}
\label{eq53} {\cal{W}}_{s}=\frac {\left(
f\,\dot{\phi}^{2}\lambda\,\sqrt {1-f\dot{\phi}^{2}}-\lambda\,\sqrt
{1-f\dot{\phi}^{2}}+1 \right) \dot{\phi}^{2}}{2{H}^{2} \left(
1-f\dot{\phi}^{2} \right) ^{3/2}}\,.
\end{equation}
In the case with $\lambda=0$, the above equations reduce to the ones
obtained in ordinary DBI setup. Although we have derived a new
${\cal{W}}_{s}$, but we can still use equations (\ref{eq19}),
(\ref{eq20}) and (\ref{eq25}) for the scalar spectral index, its
running and tensor-to-scalar ratio (where, the slow-roll parameters
and sound speed are now corresponding to the MDBI model).\\

For instance, the tensor-to-scalar ratio in the MDBI model becomes
\begin{equation}
\label{eq53c} r=16
\left(\frac{\left(1-f\dot{\phi}^{2}\right)^{-\frac{1}{2}}-2\lambda}{\left(1-f\dot{\phi}^{2}\right)
^{-\frac{3}{2}}-2\lambda}\right)^{\frac{1}{2}} \epsilon\,,
\end{equation}
or
\begin{equation}
\label{eq53c} r=-8
\left(\frac{\left(1-f\dot{\phi}^{2}\right)^{-\frac{1}{2}}-2\lambda}{\left(1-f\dot{\phi}^{2}\right)
^{-\frac{3}{2}}-2\lambda}\right)^{\frac{1}{2}} n_{T}\,.
\end{equation}
Here also, if we take $\lambda=0$ the above equations become the same as
equations (\ref{eq25}) and (\ref{eq26}).\\

Also, to seek for the non-Gaussian features of the primordial
perturbations we can use equation (\ref{eq31}), (\ref{eq32}) and
(\ref{eq33}). It is important to notice that, although our model is
the MDBI model, but the relations (\ref{eq31}), (\ref{eq32}) and
(\ref{eq33}) are applicable. This is because the MDBI model is still
$P(X,\phi)$ one and the differences appear in the definitions of the
slow-roll parameters and also the sound speed.\\

This is because for the MDBI model $P(X,\phi)$ is given by
\begin{eqnarray}
\label{eq54a}
P(X,\phi)=-f^{-1}(\phi)\sqrt{1-2Xf(\phi)}-V(\phi)+\lambda\Big(1-2X\Big)
\end{eqnarray}
which differs from $P(X,\phi)$ of the DBI model in $\lambda(1-2X)$ term.
This term is linear in $X$, so there would be no change in the
contribution of equation (\ref{eq33b}) and we end up with the
amplitude of the non-Gaussianity as given by equation (\ref{eq33}).\\

The Lagrange Multiplier $\lambda$ in equations
(\ref{eq48})-(\ref{eq53}) is not determined yet. This parameter can
be found by taking trace of the Einstein's field equation
(\ref{eq48}). However, we don't track this way. As before, following
Refs.~\cite{Bam14,Odi15}, we use the Friedmann equations to obtain
the Lagrange Multiplier and potential in terms of the Hubble
parameter and its derivatives. Here also, we adopt
$f^{-1}(\phi)=V(\phi)$. Now, by considering $\phi=t$, leading to
$\dot{\phi}=1$, equation (\ref{eq50}) gives the following expression
for the potential
\begin{eqnarray}
\label{eq54} V=\frac {\Big[2H(N) H'(N) +3H^{2}(N)
\Big]^{2}}{{\kappa}^{2} \Big[-{\kappa}^{2}+4H(N) H'(N)+6H^{2}(N)
 \Big] }\,,
\end{eqnarray}
where a prime shows a derivative of the parameter with respect to
$N$. We can obtain the Lagrange Multiplier from equations
(\ref{eq49}) and (\ref{eq54}) as follows
\begin{eqnarray}
\label{eq55} \lambda=\frac
{9\,{H}^{4}B-3\,{H}^{2}B{\kappa}^{2}-9\,{H}^{4}-
12\,{H}^{3}H'-4\,H^{2}H'^{2}B-4\,H^{2}H'^{2}}{-2AB{\kappa}
^{2}},\nonumber\\
\end{eqnarray}
where $H=H(N)$ and
\begin{equation}
\label{eq56} A=6\,{H}^{2}+4\,H\,H'-{\kappa}^ {2}  \quad
,\quad B=\sqrt {1-{\frac {{\kappa}^{2}\,A }{ \Big(
3\,{H}^{2}+2\,H\,H' \Big) ^{2}}}}\,.
\end{equation}
We also find the following expressions for the slow-roll parameters
in terms of the Hubble parameter and its derivatives
\begin{eqnarray}
\label{eq57}\epsilon=\frac{\kappa}{2}\,\sqrt{-\frac{2H'}{H^{3}}}\,,
\end{eqnarray}

\begin{eqnarray}
\label{eq58}\eta=\frac{\sqrt {2}\,\kappa}{4}\,\frac {\left( H\,
H''+H'^{2 } \right) }{ \left( -H\, H' \right) ^{\frac{3}{2}}}\,.
\end{eqnarray}
The third parameter $s$ is lengthy and has been shifted to the Appendix.
By having the slow-roll parameters, we can write the
scalar spectral index, its running and the tensor-to-scalar ratio in
terms of the Hubble parameter and its derivatives to explore them
numerically.

\subsection{Power-Law Inflation in Mimetic DBI Model}

To study the power-law inflation in MDBI model, we use the scale
factor defined in equation (\ref{eq34}), corresponding to the Hubble
parameter (\ref{eq35}). By this scale factor, the slow-roll
parameters in the MDBI model take the following form
\begin{eqnarray}
\label{eq59}\epsilon=\eta=\frac{1}{2}\,\frac{\sqrt {2}\kappa\,{{\rm
e}^{{\frac {N}{n}}}}}{{n}^{\frac{3}{2}}}\,,
\end{eqnarray}

\begin{eqnarray}
\label{eq60}s=-\frac{\sqrt {2}}{2}{{\rm e}^{-{\frac {N}{n}}}} \left(
108{{\rm e}^{-2{ \frac {N}{n}}}}{n}^{4}-216\,{{\rm e}^{-2\,{\frac
{N}{n}}}}{n}^{3}+144 \,{n}^{2}{{\rm e}^{-2\,{\frac
{N}{n}}}}-9\,{n}^{2}{\kappa}^{2}-32\,n{ {\rm e}^{-2\,{\frac
{N}{n}}}}+12\,n{\kappa}^{2}-4\,{\kappa}^{2} \right) \nonumber\\
{\kappa}^{5}{n}^{-\frac{3}{2}} \left( 3{n}^{2}{{\rm e}^{-2\,{\frac
{N}{n}}}}-2\,n{{\rm e}^{-2{\frac {N}{n}}}}-{\kappa}^{2} \right) ^{-1
} \Big( 54{{\rm e}^{-6\,{\frac {N}{n}}}}{n}^{5}-108\,{{\rm e}^{-6\,
{\frac {N}{n}}}}{n}^{4}+72\,{{\rm e}^{-6\,{\frac
{N}{n}}}}{n}^{3}-54\, {n}^{3}{{\rm e}^{-4\,{\frac
{N}{n}}}}{\kappa}^{2}\nonumber\\-16{{\rm e}^{-6{ \frac
{N}{n}}}}{n}^{2}+72{n}^{2}{{\rm e}^{-4{\frac {N}{n}}}}{
\kappa}^{2}+18{n}^{2}{{\rm e}^{-2{\frac {N}{n}}}}{\kappa}^{4}-24
n{{\rm e}^{-4{\frac {N}{n}}}}{\kappa}^{2}-6n{{\rm e}^{-2{\frac {
N}{n}}}}{\kappa}^{4}-4{{\rm e}^{-2{\frac
{N}{n}}}}{\kappa}^{4}-3 {\kappa}^{6} \Big) ^{-1}.\nonumber\\
\end{eqnarray}
By using the above relations, the scalar spectral index, its running
and tensor-to-scalar ratio can be expressed in terms of $N$ and $n$
and then explored numerically. However, we first analyze the stability
issue in the MDBI model. We study $c_{s}^{2}$ and ${\cal{W}}_{s}$
numerically to find some subspaces of the model parameters space for which
the model is free of gradient and ghost instabilities. In the left
panel of figure 5, we see the behavior of the sound speed versus $n$
in the interval $50\leq N\leq 70$. Our numerical analysis shows that for
$N=50$ the MDBI model is free of gradient instabilities and
also satisfies the causality requirement if $n\leq 14.6$. For
$N=60$, it is free of the gradient instabilities and satisfies the
causality requirement if $n\leq 16.7$. Also, for $N=70$, the
constraint on $n$ in order to get rid of gradient
instabilities is $n\leq 19$. Table 1 shows the viable ranges
of $n$ for several sample values of $N$. In the right panel of
figure 5, we see the behavior of ${\cal{W}}_{s}$ versus $n$ in the interval
$50\leq N\leq 70$. Our numerical analysis shows that for
$N=50$, in the domains $n\leq 15.3$ and $n\geq 18.2$, the parameter
${\cal{W}}_{s}$ is positive and therefore the power-law MDBI model
is free of the ghost instability. For $N=60$, in the domains $n\leq 17.6$ and
$n\geq 20.9$, and also for $N=70$, in the domains $n\leq 19.8$ and $n\geq
23.4$, the parameter ${\cal{W}}_{s}$ is positive and the model is free of
ghost instability. Table 2 shows the acceptable ranges of $n$,
leading to positive ${\cal{W}}_{s}$ for several sample values of
$N$. So, regarding the stability issue of the MDBI model,
there are some subspaces of the model parameters space
that the model is free of the ghost and gradient instabilities.

\begin{figure}[]
\begin{center}
\includegraphics[scale=0.37]{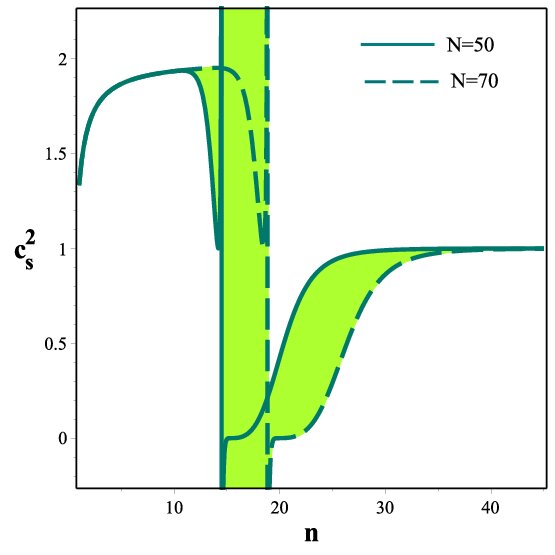}
\includegraphics[scale=0.37]{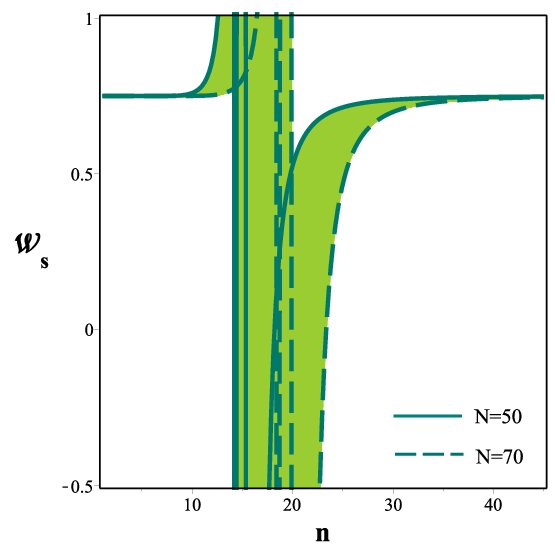}
\end{center}
\caption{\small {$c_{s}^{2}$ and ${\cal{W}}_{s}$ versus
$n$ in the MDBI inflation with power-law scale factor as
$a=a_{0}\,t^{n}$. The colored regions are free of instabilities.
Note that there is overlap between these regions in two panels.}}
\label{fig5}
\end{figure}

\begin{small}
\begin{table*}
\caption{\label{tab:1} The ranges of $n$ in which the MDBI model is
free of the gradient instabilities and also satisfies the causality
requirement. The ranges are obtained for some sample values of the
number of e-folds. }
\begin{tabular}{cccccccccc}
\\ \hline \hline $N$& $50$&&$55$&&$60$
&&$65$&&$70$\\ \hline\\
MDBI&  $14.6\leq n$ &&$15.6\leq n$&&$16.7\leq n$ &&$17.8\leq n$
&&$19\leq n$\\
\hline\hline
\end{tabular}
\end{table*}
\end{small}

\begin{small}
\begin{table*}
\caption{\label{tab:2} The ranges of $n$ in which the MDBI model is
free of the ghost instabilities. The ranges are obtained for some
sample values of the number of e-folds. }
\begin{tabular}{cccccccccc}
\\ \hline \hline $N$& $50$&&$55$&&$60$
&&$65$&&$70$\\ \hline\\
MDBI&  $n\leq 15.3$ &&$ n \leq 16.5$&&$n\leq 17.6$ &&$n\leq 18.7$
&&$n\leq 19.8$\\
 & $\&$   &&$\&$ && $\&$ && $\&$
&& $\&$\\
&  $18.2\leq n$ &&$ 19.6\leq n$&&$20.9 \leq n$ &&$22.1 \leq n$
&&$23.4 \leq n$\\
\hline\hline
\end{tabular}
\end{table*}
\end{small}

\begin{figure}[]
\begin{center}
\includegraphics[scale=0.3]{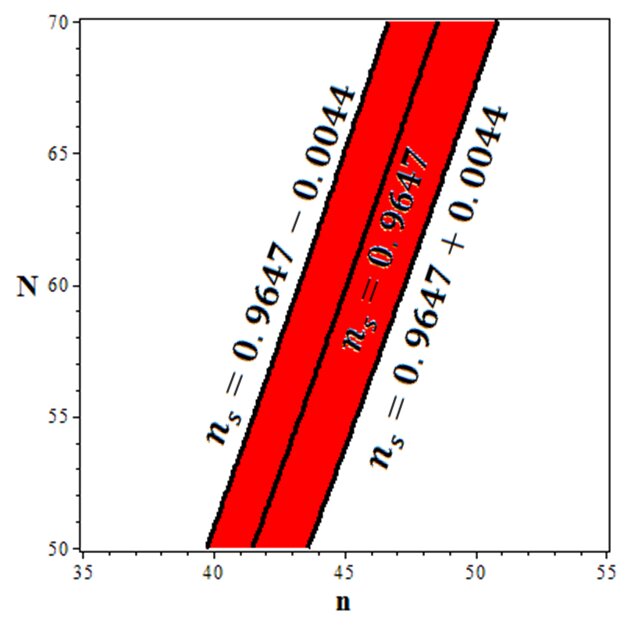}
\includegraphics[scale=0.3]{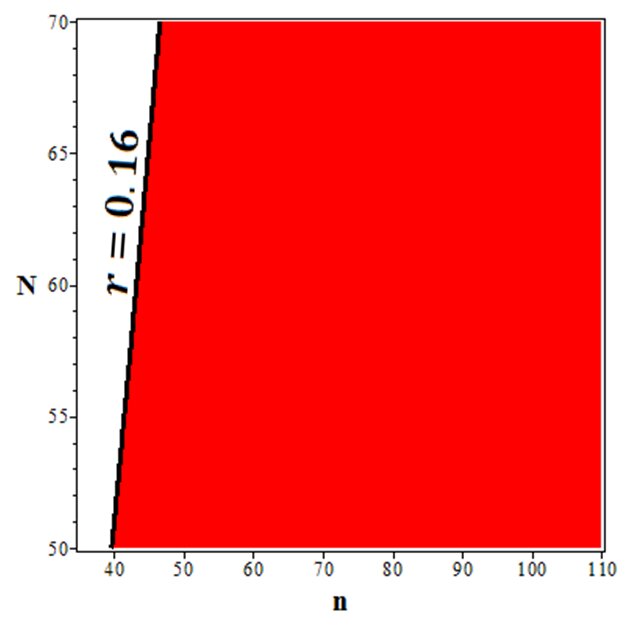}
\includegraphics[scale=0.3]{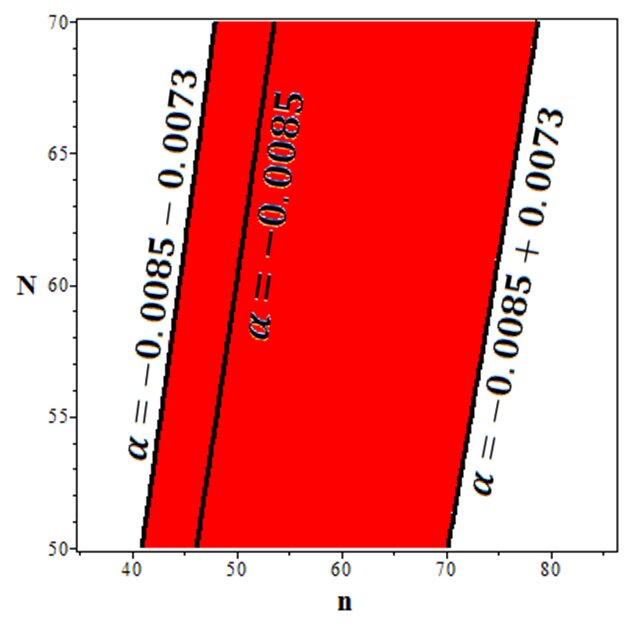}
\end{center}
\caption{\small {The ranges of parameters $n$ and $N$
which lead to the observationally viable values of the scalar
spectral index (upper-left panel), the tensor-to-scalar ratio
(upper-right panel) and the running of the scalar spectral index
(lower panel) in the MDBI model with $a=a_{0}\,t^{n}$. The adopted
values of $n_{s}$, $r$ and $\alpha_{s}$ are from Planck2018
TT, TE, EE+low EB+lensing data, in $\Lambda$CDM$+r+\frac{dn_{s}}{d\ln
k}$ model.}}
\label{fig6}
\end{figure}

Now, we find some constraints on the parameter $n$ by exploring the
observational viability of $n_{s}$, $r$ and $\alpha_{s}$, in
confrontation with the Planck2018 dataset. Firstly, we study $n_{s}$,
$r$ and $\alpha_{s}$ separately. The results are shown in figure 6.
The upper-left panel of figure 6 shows the parameters space of the
MDBI model which leads to $n_{s}=0.9647\pm 0.0044$. Our numerical
analysis shows that for $N=50$ the scalar spectral index would be
viable if $39.71\leq n \leq 43.55$. For $N=60$ we get the viable
scalar spectral index if $43.25\leq n \leq 47.29$. For $n=70$ the
scalar spectral index is viable if $46.665\leq n \leq 50.75$. The
upper-right panel of figure 6 shows the parameter space of the MDBI
model leading to the observationally viable values of the
tensor-to-scalar ratio. Based on this numerical analysis, the
tensor-to-scalar is observationally viable if $n\geq 39.21$ for
$N=50$, $n\geq 43.02$ for $N=60$ and $n\geq 46.07$ for $N=70$. The
lower panel of figure 6 shows the parameter space of the MDBI model
leading to the observationally viable values of the running of the
scalar spectral index. This numerical analysis shows that the running
of the scalar spectral index is observationally viable if $40.79\leq
n \leq 70.23$ for $N=50$, $44.51\leq n \leq 74.55$ for $N=60$ and
$47.78\leq n \leq 78.81$ for $N=70$.

Although by studying $n_{s}$, $r$ and $\alpha_{s}$ we have obtained
some constraints on the parameter $n$, however, it is useful to find
the constraints by exploring $\alpha_{s}-n_{s}$ and $r-n_{s}$
behavior in comparison to the Planck2018 data. We have performed
numerical analysis to obtain some constraints on the parameter $n$.
These constraints show the ranges of the parameters in which both
the scalar spectral index and tensor-to-scalar ratio are consistent
with Planck2018 data. Scalar spectral index versus the running of
the scalar spectral index is shown in figure 7, where we have used
the Planck2018 TT, TE, EE+lowE+lensing data at $68\%$ and $95\%$ CL
in the background. In this figure, we have plotted $n_{s}$ versus
$\alpha_{s}$ for three values of the e-folds number as $N=50, 60$
and $70$. As figure shows, $\alpha_{s}-n_{s}$ plane in the MDBI
model lies in the 68\% and 95\% CL of the Planck2018 TT, TE, EE+low
E+lensing data. By using this dataset, we have obtained some
constraints on $n$. Our numerical analysis shows that,
$\alpha_{s}-n_{s}$ in the MDBI model is observationally viable if
$38.7 \leq n\leq 47.6$ for $N=50$, $42.2\leq n\leq 51.2$ for $N=60$
and $45.6 \leq n\leq 54.9$ for $N=70$. In table 3 the constraints
obtained for some sample values of the e-folds number are presented.
Figure 8 shows the tensor-to-scalar ratio versus the scalar spectral
index in the MDBI model, in the background of the Planck2018 TT, TE,
EE+low E+lensing data at the $68\%$ and $95\%$ CL, for $N=50, 60$
and $70$. Although $r-n_{s}$ planes for all three values of the
e-folds number overlap, however, the observationally viable ranges
of $n$ are different for each cases. Our numerical analysis shows
that $r-n_{s}$ in the MDBI model is observationally viable if
$43.3\leq n\leq 46.2$ for $N=50$, $47.0 \leq n\leq 50.1$ for $N=60$
and $50.5 \leq n\leq 53.5$ for $N=70$. In table 4 the constraints
obtained for some sample values of the e-folds number are presented.
Note that, the constraints obtained from $r-n_{s}$ analysis are the
subsets of the ones obtained by analyzing $\alpha_{s}-n_{s}$.
Therefore, these constraints are more viable in the sense of leading
to observationally viable behavior of both $r-n_{s}$ and
$\alpha_{s}-n_{s}$. In this regard, we use the constraint obtained
from $r-n_{s}$ analysis to constraint the sound speed of the model
and non-Gaussian feature. In table 4, you also see the constraints
on $c_{s}$ and $f_{NL}$ corresponding to the observationally viable
values of the scalar spectral index, its running and
tensor-to-scalar ratio. Figure 9 shows the nonlinearity parameter
versus the sound speed in the MDBI model. The right panel of this
figure shows the observationally viable ranges of the nonlinearity
parameter which is obtained from the viable values of the sound
speed in this model. According to our numerical analysis, the MDBI
model with power-law scale factor is consistent with observational
data for some ranges of $n$ and predicts small amplitudes of the
non-Gaussianity.

\begin{small}
\begin{table*}
\caption{\label{tab:3} The ranges of $n$ in which both the scalar
spectral index and its running in the MDBI inflation are consistent
with 95\% CL of the Planck2018 TT, TE, EE+low E+lensing data. }
\begin{tabular}{cccccccccc}
\\ \hline \hline $N$& $50$&&$55$&&$60$
&&$65$&&$70$\\ \hline\\
&  $37.9\leq n \leq 47.2$ &&$39.7\leq n \leq 49.1$&&$41.5\leq n \leq
50.7$ &&$43.1\leq n \leq 52.6$
&&$44.8\leq n \leq 54.3$\\
\hline\hline
\end{tabular}
\end{table*}
\end{small}

\begin{figure}[]
\begin{center}
\includegraphics[scale=0.26]{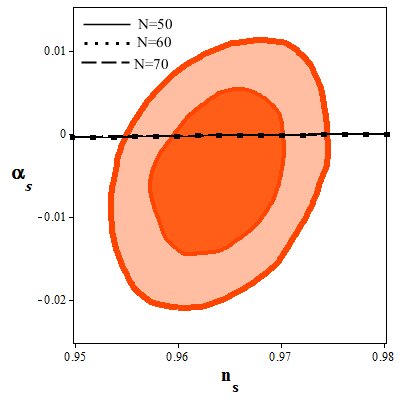}
\includegraphics[scale=0.37]{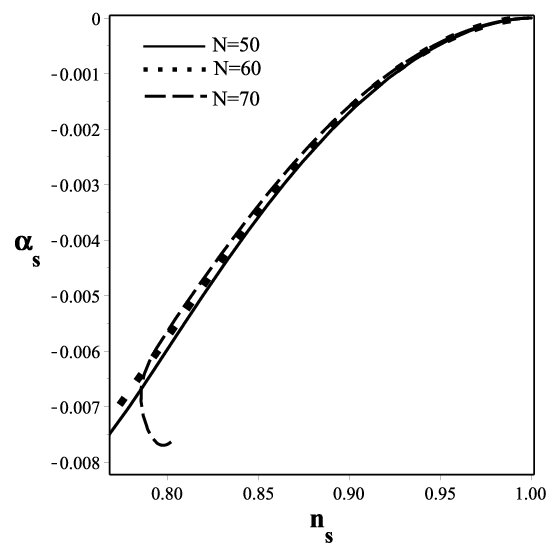}
\end{center}
\caption{\small {Running of the scalar spectral index
versus the scalar spectral index of the MDBI model in the background
of the Planck2018 TT, TE, EE+lowE+lensing data. In the right panel we
have zoomed out the evolution of $\alpha_{s}$ versus $n_{s}$.}}
\label{fig7}
\end{figure}

\begin{figure}[]
\begin{center}
\includegraphics[scale=0.3]{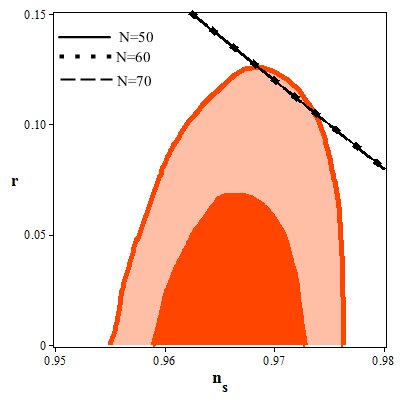}
\end{center}
\caption{\small {Tensor-to-scalar ratio versus the
scalar spectral index in MDBI model with $a=a_{0}\,t^{n}$ in the
background of Planck2018 TT, TE, EE+low E+lensing data.}}
\label{fig8}
\end{figure}

\begin{figure}[]
\begin{center}
\includegraphics[scale=0.34]{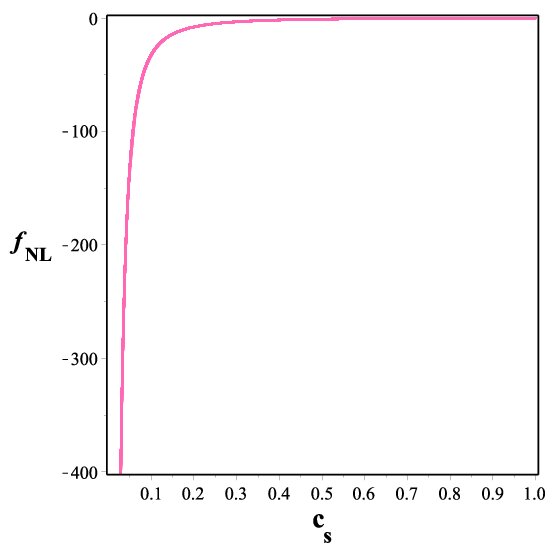}
\includegraphics[scale=0.4]{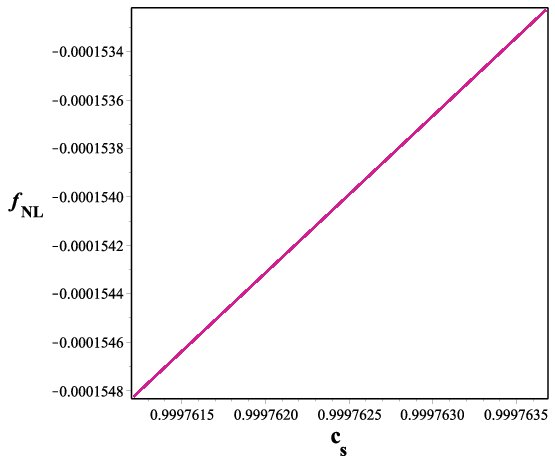}
\end{center}
\caption{\small {The nonlinearity parameter versus the
sound speed in MDBI model with $a=a_{0}\,t^{n}$ (left panel).
The observationally viable ranges of the nonlinearity parameter and
sound speed in this model, obtained from the constraints on $n$ (right panel).}}
\label{fig9}
\end{figure}

\begin{table*}
\begin{tiny}
\caption{\label{tab:4} The ranges of $n$ in which both the scalar
spectral index and the tensor-to-scalar ratio in the MDBI inflation
are consistent with 95\% CL of the Planck2018 TT, TE, EE+low E+lensing
data. The table also shows the constraints on the sound speed and
the nonlinearity parameter corresponding to the constraints on $n$.}
\begin{tabular}{ccccc}
\\ \hline \hline \\ $N=50$&$43.3\leq n\leq 46.2$&$-0.20214\times 10^{-3}\leq
\alpha_{s}\leq -0.14817\times 10^{-3} $&$0.99978\leq
c_{s}\leq 0.99987$\\
\\
&&&$-0.13764\times 10^{-3}\leq f_{NL} \leq -0.846199\times 10^{-4}$\\
\hline \\
$N=55$&$45.1\leq n\leq 48.1$&$-0.19870\times 10^{-3}\leq
\alpha_{s}\leq-0.14482\times 10^{-3} $
&$0.99976\leq c_{s} \leq 0.99985$\\
\\
&&
&$-0.15777\times 10^{-3}\leq f_{NL} \leq-0.95739\times 10^{-4} $\\
\hline \\ $N=60$&$47.0 \leq n\leq 50.1$&$-0.19273\times 10^{-3} \leq
\alpha_{s}\leq -0.13989\times 10^{-3}$ &$ 0.99973\leq
c_{s}\leq 0.99984$\\
\\ &&
&$ -0.17508\times 10^{-3}\leq f_{NL} \leq-0.10517\times 10^{-3} $\\
\hline \\ $N=65$&$48.8 \leq n\leq 52.0$&$-0.18862\times 10^{-3} \leq
\alpha_{s}\leq -0.13626\times 10^{-3}$ &$0.99969\leq c_{s}\leq
0.99982 $\\ \\&&
&$-0.19512\times 10^{-3}\leq f_{NL} \leq -0.11594\times 10^{-3}$\\
\hline \\ $N=70$&$50.5 \leq n\leq 53.5$&$-0.15809\times 10^{-3} \leq
\alpha_{s}\leq -0.11817\times 10^{-4}$ &$0.99966 \leq
c_{s}\leq 0.99979$\\
\\ &&
&$-0.21862\times 10^{-3}\leq f_{NL} \leq -0.13446\times 10^{-3}$\\
\hline \hline
\end{tabular}
\end{tiny}
\end{table*}

\section{Constant Sound Speed}

In this section, we consider the case in which the sound speed is
constant and study the DBI and MDBI models with constant $c_{s}$.
In this section, we use the Planck constraint on $f_{NL}$ and
$c_{s}$ to obtain the observationally viable values of $n$. In fact,
for some constant values of the sound speed- which is corresponding
to the observationally viable values of nonlinearity parameter- we
study $r-n_{s}$ and $\alpha_{s}-n_{s}$ and find some constraints on
$n$. The background equations are the same as the previous sections
but now with constant sound speed. However, the potential and Lagrange
multiplier change and therefore some parameters such as the
slow-roll parameters, scalar spectral index and tensor-to-scalar
ratio change accordingly. We start the issue with DBI model.

\subsection{DBI model}

To obtain the DBI parameter $f$ in terms of the sound speed, we use equation
(\ref{eq10}). The result is as follows
\begin{eqnarray}
\label{eq61} f=\frac{1}{2}\frac {{\kappa}^{2} \left( c_{s}^{2}-1
\right) }{H  H'}\,.
\end{eqnarray}
By substituting this expression for $f$ in equation (\ref{eq6}), we
find the potential of the model as
\begin{eqnarray}
\label{eq62} V=-{\frac {H \left( -3\,c_{s}^{3}\,H +3\,H \,
c_{s}+2\,H' \right) }{{\kappa }^{2} \left( c_{s}^{2}-1 \right)
c_{s}}}\,.
\end{eqnarray}
Now, by having the functions $f$ and $V$, we use equations
(\ref{eq6}), (\ref{eq9})  and (\ref{eq34}) to find the following
slow-roll parameters
\begin{eqnarray}
\label{eq63} \epsilon={\frac
{9c_{s}^{6}{n}^{2}-18c_{s}^{4}{n}^{2}+12n
c_{s}^{3}-12c_{s}^{4}n+9c_{s}^{2}{n}^{2}-12nc_{s}+ 12c_{s}^{2}n-8
c_{s}+4 c_{s}^{2}+4}{n \left( 3c_{s}^{3}n-3c_{s}\,n+2-2\,c_{s}
\right) ^{2}}}\,,
\end{eqnarray}

\begin{eqnarray}
\label{eq64} \eta=-\frac{1}{2}\Bigg( 6c_{s}^{6}n{{\rm e}^{-4{\frac
{N}{n}}}}{\kappa} ^{2}-9c_{s}^{6}{n}^{3}{{\rm e}^{-6{\frac
{N}{n}}}}-12c_{s}^{4}n{{\rm e}^{-4{\frac
{N}{n}}}}{\kappa}^{2}+18c_{s}^{4} {n}^{3}{{\rm e}^{-6{\frac
{N}{n}}}}+4{{\rm e}^{-4{\frac {N}{n}}} } c_{s}^{3}{\kappa}^{2}+4{
{\rm e}^{-4{\frac
{N}{n}}}}c_{s}^{4}{\kappa}^{2}\nonumber\\-12{n}^{2}{ {\rm
e}^{-6{\frac {N}{n}}}} c_{s}^{3}+12c_{s}^{4}{n}^{2}{{\rm
e}^{-6{\frac {N}{n}}}}+6c_{s}^{2}n{{\rm e}^{-4{\frac
{N}{n}}}}{\kappa}^{2}-9c_{s}^ {2}{n}^{3}{{\rm e}^{-6{\frac
{N}{n}}}}-4{{\rm e}^{-4{\frac {N}{n }}}}c_{s}{\kappa}^{2}-4{ {\rm
e}^{-4{\frac {N}{n}}}}c_{s}^{2}{\kappa}^{2}\nonumber\\+12{n}^{2}{
{\rm e}^{-6{\frac {N}{n}}}} c_{s}-12c_{s}^{2}{n}^{2}{{\rm
e}^{-6{\frac {N}{n}}}}+8n{ {\rm e}^{-6{\frac {N}{n}}}} c_{s}-4n{{\rm
e}^{-6{\frac {N}{n}}}}c_{s}^{2}-4n{{\rm e}^{-6 {\frac {N}{n}}}}
\Bigg) \frac{{{\rm e}^{6{\frac {N}{n}}}}{n}^{-2}}{ \bigg(
3c_{s}^{3}n -3c_{s}n+2 \bigg) ^{2}}.
\end{eqnarray}
In this case also, the third slow-roll parameter, $s$, is zero.
After obtaining the slow-roll parameters, we can find scalar
spectral index, its running and tensor-to-scalar ratio and then analyze
the parameters space of the model numerically. Since the sound speed is constant in this section,
we can use Planck constraint on this quantity to analyze the model numerically. We adopt some sample values of the sound
speed which are larger than $0.087$ (the lower limit from Planck2015
temperature and polarization data at 95\% CL~\cite{pl15}). Note
that, although $\eta$ is related to the number of e-folds parameter,
but it has not an impressive effect on the numerical analysis of the model. In
fact, the difference between the values of $n_{s}$ and $r$ for
different values of $N$ is of the order of $10^{-7}$. Therefore, the
constraint that we obtain, is very nearly independent of the values
of the e-folds number. We perform a numerical analysis on the
model's parameters for some sample values of the sound speed as
$c_{s}=0.1$, $0.4$, $0.7$ and $0.9$. By using these values of
$c_{s}$, we study $r-n_{s}$ plane in comparison with Planck2018
TT, TE, EE+lowE+lensing data. The results are shown in figure 10. As
this figure shows, the DBI model with constant sound speed and the
power-law scale factor, in some ranges of $n$ is consistent with
Planck2018 observational data. The ranges of $n$ leading to the
observationally viable values of the scalar spectral index and
tensor-to-scalar ratio are shown in tables 5 and 6. Then, we study
$\alpha_{s}-n_{s}$ plane in comparison with Planck2018
TT, TE, EE+lowE+lensing data, as shown in figure 11. In this regard, we
obtain the values of the running of the scalar spectral index
corresponding to these ranges of $n$. The results are shown in
tables 5 and 6.

Note that, the power-law DBI model, with $f$ and $V$ as given in
equations (\ref{eq61}) and (\ref{eq62}) is ghost free. This is
because that in this case, the parameter ${\cal{W}}$, given by
(\ref{eq18}), is always positive for values of $n\geq 8$ and
$c_{s}\geq 0.087$.

\begin{figure}[]
\begin{center}
\includegraphics[scale=0.3]{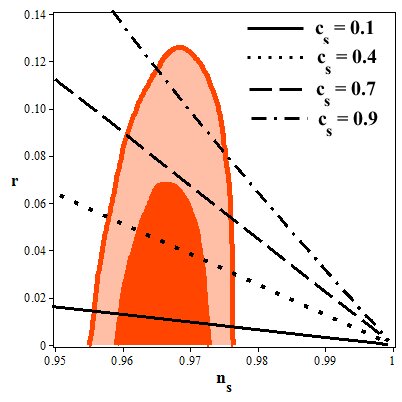}
\end{center}
\caption{\small {Tensor-to-scalar ratio versus the
scalar spectral index for the power-law DBI inflation with constant
sound speed in the background of Planck2018 TT, TE, EE+lowE+lensing
data.}}
\label{fig10}
\end{figure}

\begin{figure}[]
\begin{center}
\includegraphics[scale=0.29]{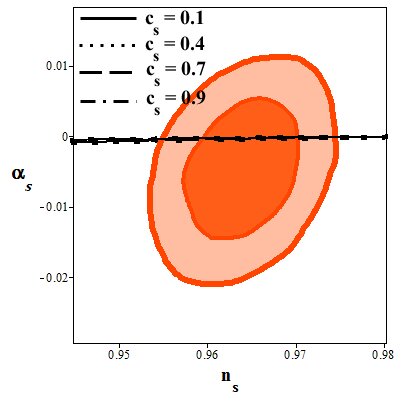}
\includegraphics[scale=0.4]{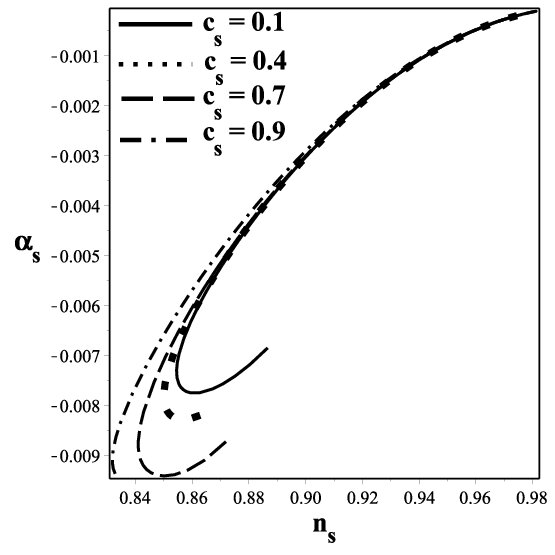}
\end{center}
\caption{\small {Running of the scalar spectral index
versus the scalar spectral index for the power-law DBI inflation
with constant sound speed in the background of the Planck2018
TT, TE, EE+lowE+lensing data. In the right panel we have zoomed out
the evolution of $\alpha_{s}$ versus $n_{s}$.}}
\label{fig11}
\end{figure}

\begin{table*}
\begin{small}
\caption{\small{\label{tab:5} The ranges of $n$ in which both
the scalar spectral index and tensor-to-scalar ratio in the
power-law DBI inflation with constant sound speed are consistent
with 95\% CL of the Planck2018 TT, TE, EE+lowE+lensing data. This
table also shows the constraints on the running of the scalar
spectral index obtained from the constraints on $n$.}}
\begin{tabular}{ccccc}
\\ \hline \hline \\ $c_{s}$&$f_{NL}$&$n$&$\alpha_{s}$\\
\hline \\ $0.1$&$-32.08333$&$112.7\leq
n\leq 211.1$&$-0.58770\times10^{-3}\leq \alpha_{s} \leq -0.16783\times 10^{-3}$\\
\hline \\ $0.4$&$-1.70139$&$118.5\leq n\leq 210.3$
&$-0.53110\times10^{-3}\leq \alpha_{s} \leq -0.16899\times 10^{-3}$\\
\hline \\ $0.7$&$-0.33730$&$126.3\leq n\leq 208.7$
&$-0.46504\times10^{-3}\leq \alpha_{s} \leq -0.17102\times 10^{-3}$\\
\hline \\ $0.9$&$-0.07602$&$134.1\leq n\leq 204.8$
&$-0.40302\times10^{-3}\leq \alpha_{s} \leq -0.17490\times 10^{-3}$\\
\hline \hline
\end{tabular}
\end{small}
\end{table*}

\begin{table*}
\begin{small}
\caption{\small{\label{tab:6} The ranges of $n$ in which both
the scalar spectral index and tensor-to-scalar ratio in the
power-law DBI inflation with constant sound speed are consistent
with 68\% CL of the Planck2018 TT, TE, EE+lowE+lensing data. This
table also shows the constraints on the running of the scalar
spectral index obtained from the constraints on $n$.}}
\begin{tabular}{ccccc}
\\ \hline \hline \\ $c_{s}$&$f_{NL}$&$n$&$\alpha_{s}$\\
\hline \\ $0.1$&$-32.08333$&$123.4\leq
n\leq 181.8$&$-0.49035\times10^{-3}\leq \alpha_{s} \leq -0.22620\times 10^{-3}$\\
\hline \\ $0.4$&$-1.70139$&$132.4\leq n\leq 176.5$
&$-0.42564\times10^{-3}\leq \alpha_{s} \leq -0.23978\times 10^{-3}$\\
\hline \\ $0.7$& not consistent& not consistent &not consistent\\
\hline \\ $0.9$&not consistent&not consistent
&not consistent\\
\hline \hline
\end{tabular}
\end{small}
\end{table*}

\subsection{Mimetic DBI (MDBI) model}

By using equations (\ref{eq49}) and (\ref{eq50}), we obtain the
lagrange multiplier in terms of the brane tension as follows
\begin{eqnarray}
\label{eq65} \lambda=\frac{1}{2}{\frac {2\,H'\,\sqrt
{1-f}+{\kappa}^{2}}{\sqrt {1-f }{\kappa}^{2}}}\,.
\end{eqnarray}
Now, by using the above equation and equation (\ref{eq52}) we find
\begin{eqnarray}
\label{eq66} f=- \Bigg[ \frac{1}{6}{\frac {\sqrt
[3]{{\kappa}^{2}c_{s}^{2} \left( -{ {\it
cs}}^{4}\,{\kappa}^{4}+54\,H'^{2}\,c_{s}^{4}-108\,H'^{2}\,c_{s}^{2}+c_{s}^{2}z+54\,H'^{2}-z
\right) }} {H'\, \left( c_{s}^{2}-1 \right) }}-\frac{1}{6}{\frac
{{\kappa}^{2}\,c_{s}^{2}}{H'\, \left( c_{s}^{2}-1 \right)
}}\nonumber\\+\frac{1}{6}{\frac {c_{s}^{4}{\kappa}^{4}}{H'\, \left(
c_{s}^{2}-1 \right) \sqrt [3] {{\kappa}^{2}c_{s}^{2} \left(
-c_{s}^{4}\,{\kappa}^{4}+54\,H'^{2}c_{s}^{4}-108\,H'^{2}{{\it
cs}}^{2}+c_{s}^{2}\,z+54\,H'^{2}-z \right) }}} \Bigg] ^{ 2}+1\,,
\end{eqnarray}
where
\begin{eqnarray}
\label{eq67} z=6\,H'\,\sqrt
{-c_{s}^{4}{\kappa}^{4}+27\,H'^{2}c_{s}^{4}-54\,H'^{2}c_{s}^{2}+27\,H'^{2}}
\sqrt {3}\,.
\end{eqnarray}
By using equation (\ref{eq50}), we obtain the potential as follows
\begin{eqnarray}
\label{eq68} V=-{\frac {{\kappa}^{2}\sqrt {1-f}- \left( 2\,H \,
H'+3\, H^{2} \right) f}{f{\kappa}^{ 2}}}\,,
\end{eqnarray}
where $f$ is given by equation (\ref{eq66}). Here, by adopting the
scale factor (\ref{eq34}), we firstly study possible existence of ghost instability in
this model. By substituting equations (\ref{eq65}), (\ref{eq66}) and
(\ref{eq68}) in equation (\ref{eq53}), we explore ${\cal{W}}_{s}$
numerically. The results are shown in figure 12, where we have used
$50\leq N \leq 70$ and $c_{s}=0.1,0.4,0.7,0.9$. Our numerical
analysis shows that, for all values of $n\geq 5$ the parameter
${\cal{W}}_{s}$ is positive and therefore the power-law MDBI model
with constant sound speed is free of ghost instability (in fact, for
$n<5$, the parameter ${\cal{W}}_{s}$ becomes imaginary).

\begin{figure}[]
\begin{center}
\includegraphics[scale=0.37]{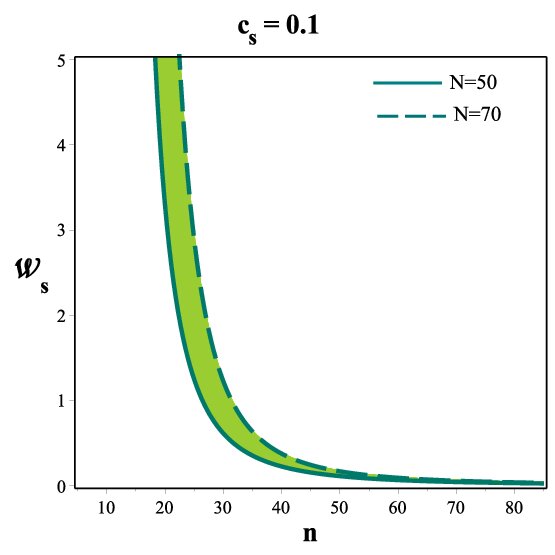}
\includegraphics[scale=0.37]{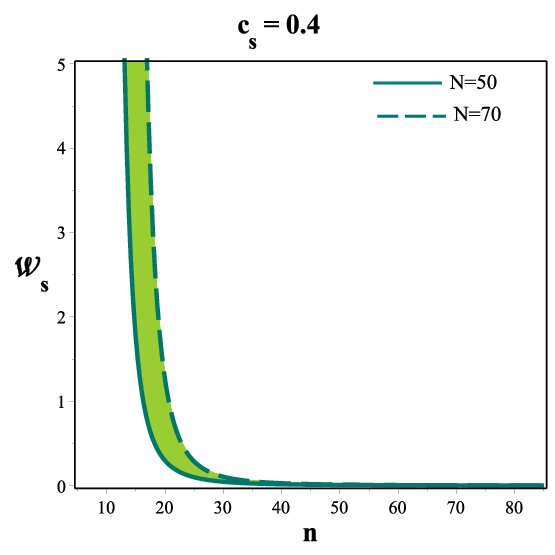}
\includegraphics[scale=0.37]{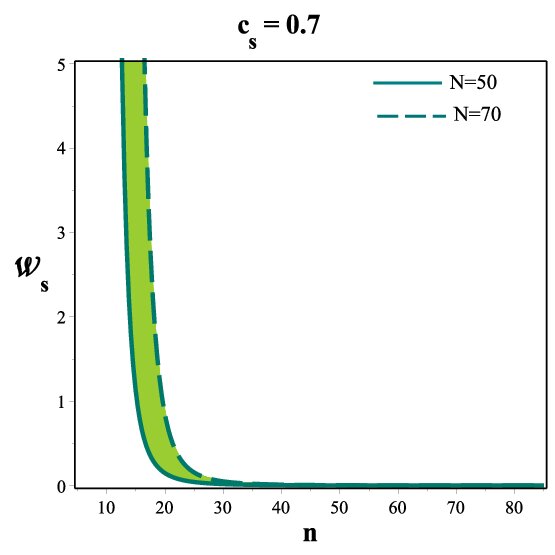}
\includegraphics[scale=0.37]{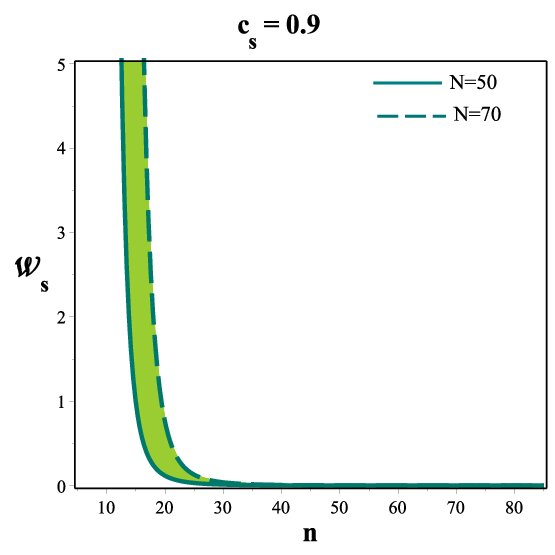}
\end{center}
\caption{\small {Evolution of ${\cal{W}}_{s}$ versus
$n$ in the domain $50\leq N\leq 70$ for power-law MDBI inflation
with constant sound speed.}}
\label{fig12}
\end{figure}

In this case, the slow-roll parameters take the following form
\begin{eqnarray}
\label{eq69} \epsilon=\frac{1}{2}\,\frac{\sqrt {6}\kappa\, \left(
3\,{{\rm e}^{-{\frac {N}{n}}}}{ n}^{2}-2\,{{\rm e}^{-{\frac
{N}{n}}}}n+1 \right)  {{\rm e}^{{\frac {N}{2n}}}}}{\sqrt {n} \left(
3\,{{\rm e}^{-{\frac {N}{n}}}}{n}^{2}-2\,{{\rm e}^{-{\frac {N}
{n}}}}n+2 \right) ^{-\frac{3}{2}}}\,,
\end{eqnarray}

\begin{eqnarray}
\label{eq70} \eta=-6\, \Bigg\{ \frac{\sqrt {3}}{6}{\kappa}^{2}
\Bigg[\frac{n}{2}\left( -\frac{8}{n} \left( {{\rm e}^{-{\frac
{N}{n}}}} \right) ^{2}+12\left( { {\rm e}^{-{\frac {N}{n}}}} \right)
^{2} \right) -\frac{1}{2}{\frac {{\kappa} ^{2}nf''}{ f^{2}\sqrt
{1-f}}}+{\frac {{\kappa}^{2}nf'^{2}}{ f^{3}\sqrt
{1-f}}}-\frac{1}{2}{ \frac {{\kappa}^{2}nf'^{2}}{ f^{2} \left(
1-f\right) ^{\frac{3}{2}}}}\nonumber\\+\frac{1}{4}{ \frac
{{\kappa}^{2}nf''}{f \left( 1-f \right)
^{\frac{3}{2}}}}+\frac{3}{8}{\frac {{\kappa}^{2}nf'^{2}}{f  \left(
1-f \right) ^{\frac{5}{2}}}}-{ \kappa}^{2}n\lambda' \Bigg] \Bigg[
\Bigg( - \bigg( {\kappa}^{2}\sqrt {1-f }- \Big( -2n \left( {{\rm
e}^{-{\frac {N}{n}}}} \right) ^{2}+{ \frac {1}{f \sqrt {1-f }}}-2
\lambda\nonumber\\+3{n}^{2} \left( { {\rm e}^{-{\frac {N}{n}}}}
\right) ^{2} \Big) f \bigg) f^{-1} \Bigg)
\Bigg]^{-\frac{1}{2}}-\frac{\sqrt {3}}{12}{\kappa}^{4} \Bigg[
\frac{\sqrt {2}}{2} \sqrt {\frac{n}{{\kappa}^{2}}} \left( 4 \left(
{{\rm e}^{-{\frac {N}{n}}}} \right) ^{2}-6n \left( {{\rm e}^{-{\frac
{N}{n}}}} \right) ^{2} \right) -\frac{1}{2}{\frac {\sqrt {2}\sqrt
{{\kappa}^{2}n}f'}{ f^{2}\sqrt {1-f}}}\nonumber\\+\frac{1}{4}{\frac
{\sqrt {2}\sqrt {{\kappa}^{2}n}f'}{f   \left( 1-f  \right) ^{3/2}}}-
\sqrt {2}\sqrt {{\kappa}^{2}n}\lambda' \Bigg] ^{2} \Bigg[ {\kappa}^{
2} \bigg( - \bigg( {\kappa}^{2}\sqrt {1-f }+ \left( 2n \left( {{\rm
e}^{-{\frac {N}{n}}}} \right) ^{2}-3{n}^{2} \left( {{\rm e}^{-{\frac
{N}{n}}}} \right) ^{2} \right) f \bigg)\frac{f^{-1}}{{\kappa}^{2}}\nonumber\\
+{\frac {1}{f \sqrt {1-f }}}-2\, \lambda \bigg) \Bigg]
^{-\frac{3}{2}} \Bigg\} {\kappa}^{-2} \Bigg( \frac{\sqrt
{2}}{2}\sqrt {{\kappa}^{2}n} \left( 4 \left( {{\rm e}^{-{\frac {N}
{n}}}} \right) ^{2}-6n \left( {{\rm e}^{-{\frac {N}{n}}}} \right) ^{
2} \right) {\kappa}^{-2}\nonumber\\-\frac{1}{2}{\frac {\sqrt
{2}\sqrt {{\kappa}^{2}n}f'}{ f^{2}\sqrt {1-f }}}+\frac{1}{4}{\frac
{\sqrt {2}\sqrt {{\kappa}^{2}n}f'}{f  \left( 1-f \right) ^{3/2}}}-
\sqrt {2}\sqrt {{\kappa}^{2}n}\lambda' \Bigg) ^{-1}\,,
\end{eqnarray}

\begin{eqnarray}
\label{eq71} s=\frac{\sqrt {6}}{8}{\frac {1}{ \sqrt {n}}} {\frac
{{\kappa}^{3}\left( -ff'\,n+2\,{f}^{2}- 2\,f\,n-2\,f \right) {{\rm
e}^{{\frac {N}{2n}}}}}{(f-1)\sqrt {3\,{{\rm e}^{-{ \frac
{N}{n}}}}{n}^{2}-2\,n{{\rm e}^{-{\frac {N}{n}}}}+2}\left( 2\, f\sqrt
{1-f}{{\rm e}^{-{\frac {N}{n}}}}-f{\kappa}^{2}-2\,{{\rm e}^{-{ \frac
{N}{n}}}}\sqrt {1-f} \right)}}\,.
\end{eqnarray}

As before, by using the obtained slow-roll parameters, we can write
the scalar spectral index, its running and tensor-to-scalar ratio in
terms of the model's parameters and analyze them numerically. In
this case, the slow-roll parameters depend on the number of e-folds.
Therefore, to obtain the observational constraints we should specify
the values of e-fold's number. We adopt three values as $N=50$, $60$
and $70$. Firstly, by using equations (\ref{eq19}) and (\ref{eq25})
(where the slow-roll parameters are given by equations
(\ref{eq69})-(\ref{eq71})), we obtain the parameter space of $c_{s}$
and $n$ leading to the observationally viable values of $r-n_{s}$
in the power-law MDBI model. To this end, we have used the
Planck2018 TT, TE, EE+lowE+lensing constraints at 68\% CL and 95\% CL.
The results are shown in figure 13, in which the dark magenta region is
corresponding to 68\% CL and the light magenta region is corresponding to
95\% CL of the Planck2018 TT, TE, EE+lowE+lensing data.

\begin{figure}[]
\begin{center}
\includegraphics[scale=0.29]{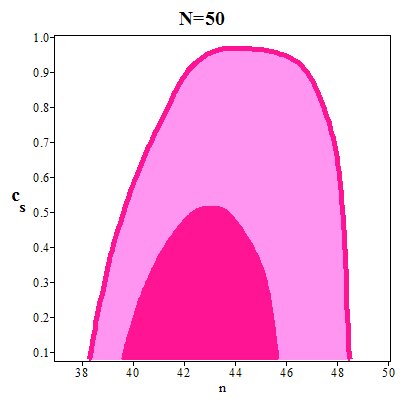}
\includegraphics[scale=0.29]{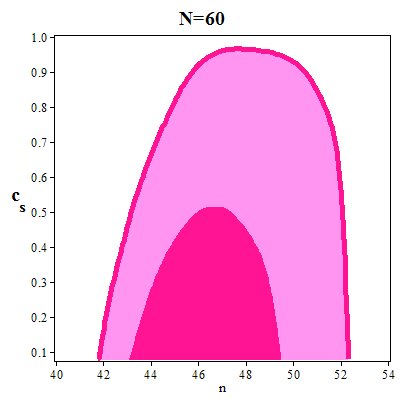}
\includegraphics[scale=0.29]{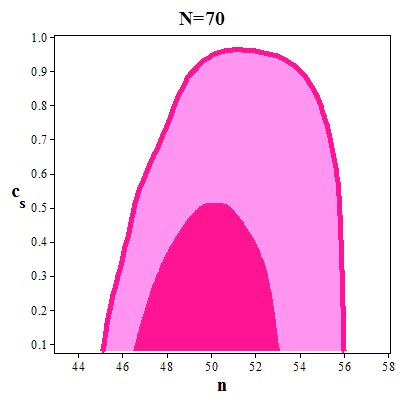}
\end{center}
\caption{\small {Ranges of parameters $c_{s}$ and $n$
in which the scalar spectral index and tensor-to-scalar ratio in a
power-law MDBI inflation with constant sound speed fulfill the
constraints obtained from Planck2018 TT, TE, EE+lowE+lensing at 68\%
CL (dark magenta region) and 95\% CL (light magenta region).}}
\label{fig13}
\end{figure}

Also, to obtain more specific numerical constraints on $n$ and use
them in order to explore the running of the scalar spectral index, we study
$r-n_{s}$ plane of the model in the background of Planck2018
TT, TE, EE+lowE+lensing. In this regard, we use the previously
adopted values of the sound speed (which are consistent with the
constraint obtained from Planck2015 temperature and polarization
data). Figure 14 shows the tensor-to-scalar ratio versus the scalar
spectral index for power-law MDBI model. By a numerical analysis we
have obtained some constraints on $n$ in which $r-n_{s}$ plane lies
within the regions of 68\% CL and 95\% CL of the Planck2018
TT, TE, EE+lowE+lensing data. These constraints are presented in
tables 7 and 8. Here also, by using the obtained constraints on $n$
and observationally viable values of $n_{s}$, we explore the running
of the scalar spectral index and find the observationally viable
values of it. The results are shown in the last column of tables 7
and 8. Also, in figure 15 the evolution of the running
of the scalar spectral index versus the scalar spectral index in the
background of Planck2018 TT, TE, EE+lowE+lensing dataset is presented.  As figure
shows, for some values of $n$, $\alpha_{s}-n_{s}$ is consistent with
observational data.

Note that the results of the amplitude of the non-Gaussianity are
the same for both MDBI and DBI models. This is because for MDBI
model also, we use equation (\ref{eq33}). Generally, the smaller
values of $c_s$ lead to the larger values of $f_{NL}$. The reason that
for the constant sound speed in DBI and MDBI models, the
non-Gaussianity is not very large is related to the form of
$P(X,\phi)$ for both DBI and MDBI models. In fact, considering that
the mimetic constraint is linear in $X$, allows us to use equation
(\ref{eq33}) which leads to the observationally viable values of the
non-Gaussianity. \\

At this stage, the difference between the obtained results for a varying sound speed and a constant sound speed needs to be more clarified.
In the DBI model with $V_{+}$, the varying sound speed is relatively large, leading to relatively small non-Gaussianity and consistent
with observational data. However, in this case the tensor-to-scalar ratio is large and not
consistent with observation. So this case is ruled out by observation. In the DBI model with $V_{-}$, the varying sound
speed is small and the amplitude of the non-Gaussianity is too large to be consistent with the observational data. This case is also ruled out by observation.
When we consider the MDBI model, the effect of the mimetic field is that it reduces the value of the tensor-to-scalar ratio.
In this regard, $r-n_{s}$ plane in the MDBI model in some ranges of the parameter $n$ becomes consistent with
Planck2018 observational data. For the observationally viable ranges of $n$, the sound speed
is large and the amplitude of the non-Gaussianity is small, which is consistent with observation.
For the varying sound speed case, we have considered $c_{s}$ as a function of the model's parameters.
In this regard, the amplitude of the non-Gaussianity also becomes a function of the model's parameters.
To constrain $c_{s}$ and $f_{NL}$ we firstly constrained the parameters of the model in confrontation
with observational data. In this regard, we have compared $r-n_{s}$ values in the MDBI model
with Planck2018 data and found some constraints on the parameter $n$.
Since $c_{s}$ and $f_{NL}$ are functions of $n$, with obtained constraints on $n$
we were able to constrain the sound speed and the amplitude of the non-Gaussianity in MDBI model.\\
In the case with a constant sound speed we have considered the sound speed as a free constant parameter.
Then the amplitude of the non-Gaussianity would be a constant corresponding to the value of $c_{s}$.
In this case, both DBI and MDBI models have the same $f_{NL}$. In the constant sound speed case,
we have obtained the potential $V$ and the DBI function $f$ (and also the Lagrange Multiplier
$\lambda$ for the MDBI model) in terms of the sound speed and other model's parameters.
Then we have used $V$, $f$ and $\lambda$ to obtain the inflation and perturbation parameters
such as $\epsilon$, $\eta$, $s$, $n_s$, $\alpha_s$ and $r$. We have adopted some
sample values of $c_s$ and by probing the observational viability of $r-n_{s}$,
we have found some constraints on the parameter $n$. Then, we have used the
observationally viable ranges of $n$ to obtain the constraints on $\alpha_s$.
Note that the dependence of the potential, DBI function and Lagrange multiplier on $c_s$
is different for the DBI and MDBI models, so the constraints on $n$ and $\alpha_s$ are different
in these two models. These are the reasons that why the results in the two sections 2.4 and 3.2 are different.\\

In summary, our proposed MDBI model with constant sound speed in some ranges
of the model's parameter space is ghost-free and consistent with
Planck2018 observational data.
We shall compare the DBI and MDBI models in details at the end of the paper.

\begin{figure}[]
\begin{center}
\includegraphics[scale=0.3]{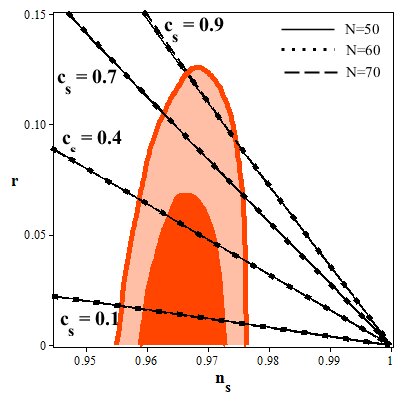}
\end{center}
\caption{\small {Tensor-to-scalar ratio versus the
scalar spectral index for the power-law MDBI inflation with
a constant sound speed in the background of Planck2018
TT, TE, EE+lowE+lensing data.}}
\label{fig14}
\end{figure}

\begin{figure}[]
\begin{center}
\includegraphics[scale=0.27]{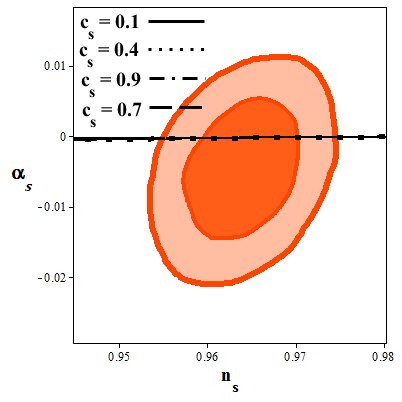}
\includegraphics[scale=0.4]{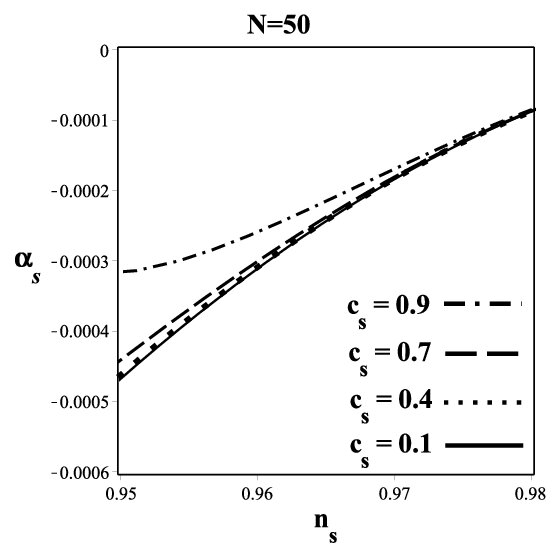}
\includegraphics[scale=0.4]{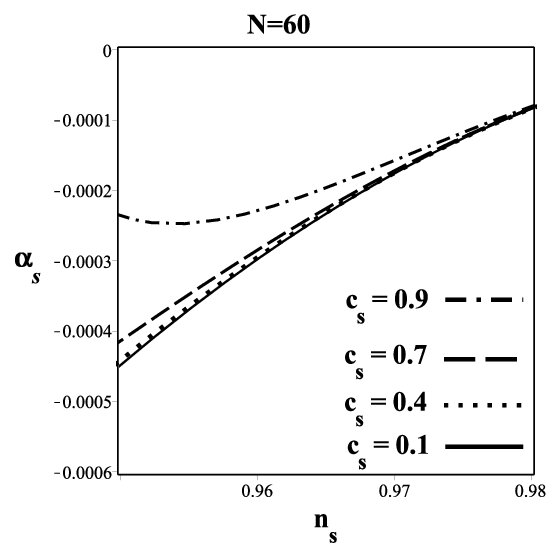}
\includegraphics[scale=0.4]{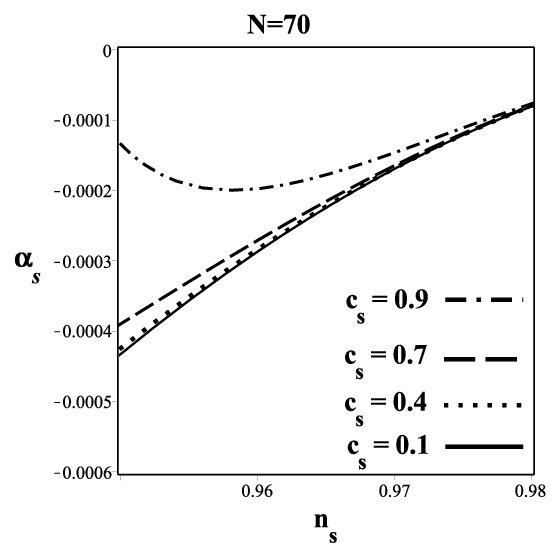}
\end{center}
\caption{\small {Running of the scalar spectral index
versus the scalar spectral index for the power-law MDBI inflation
with a constant sound speed in the background of the Planck2018
TT, TE, EE+lowE+lensing data.}}
\label{fig15}
\end{figure}

\begin{table*}
\begin{small}
\caption{\small{\label{tab:7} The ranges of $n$ in which both the
scalar spectral index and tensor-to-scalar ratio in the power-law
MDBI inflation with constant sound speed are consistent with
95\% CL of the Planck2018 TT, TE, EE+lowE+lensing data. This table
also shows the constraints on the running of the scalar spectral
index obtained from the constraints on $n$.}}
\begin{tabular}{cccccc}
\\ \hline \hline \\$N$& $c_{s}$&&$n$&$\alpha_{s}$\\
\hline \\50& $0.1$&&$38.35\leq
n\leq 48.45$&$-0.37230\times10^{-3}\leq \alpha_{s} \leq -0.11930\times 10^{-3}$\\
\hline \\
50 &$0.4$&&$39.20\leq n\leq 48.31$
&$-0.32868\times10^{-3}\leq \alpha_{s} \leq -0.12052\times 10^{-3}$\\
\hline \\
50& $0.7$&&$40.7\leq n\leq 47.9$
&$-0.26408\times10^{-3}\leq \alpha_{s} \leq -0.12353\times 10^{-3}$\\
\hline \\
50&$0.9$&&$42.0\leq n\leq46.9$
&$-0.22348\times10^{-3}\leq \alpha_{s} \leq -0.13480\times 10^{-3}$\\
\hline \hline\\
55& $0.1$&&$40.1\leq
n\leq 50.5$&$-0.36395\times10^{-3}\leq \alpha_{s} \leq -0.11523\times 10^{-3}$\\
\hline \\
55 &$0.4$&&$41.1\leq n\leq 50.3$
&$0.31643\times10^{-3}\leq \alpha_{s} \leq -0.11698\times 10^{-3}$\\
\hline \\
55& $0.7$&&$42.5\leq n\leq 49.9$
&$-0.25662\times10^{-3}\leq \alpha_{s} \leq -0.11977\times 10^{-3}$\\
\hline \\
55&$0.9$&&$43.8\leq n\leq48.8$
&$-0.21779\times10^{-3}\leq \alpha_{s} \leq -0.13161\times 10^{-3}$\\
\hline \hline\\
60& $0.1$&&$41.84\leq
n\leq 52.3$&$-0.35576\times10^{-3}\leq \alpha_{s} \leq -0.11396\times 10^{-3}$\\
\hline \\
60 &$0.4$&&$42.8\leq n\leq 52.15$
&$-0.31063\times10^{-3}\leq \alpha_{s} \leq -0.11488\times 10^{-3}$\\
\hline \\
60& $0.7$&&$44.2\leq n\leq 51.8$
&$-0.25180\times10^{-3}\leq \alpha_{s} \leq -0.11689\times 10^{-3}$\\
\hline \\
60&$0.9$&&$45.6\leq n\leq50.6$
&$-0.21207\times10^{-3}\leq \alpha_{s} \leq -0.12956\times 10^{-3}$\\
\hline \hline\\
65& $0.1$&&$43.5\leq
n\leq54.2$&$-0.34997\times10^{-3}\leq \alpha_{s} \leq -0.11128\times 10^{-3}$\\
\hline \\
65 &$0.4$&&$44.5\leq n\leq 54.0$
&$-0.30429\times10^{-3}\leq \alpha_{s} \leq -0.11265\times 10^{-3}$\\
\hline \\
65& $0.7$&&$46.0\leq n\leq 53.6$
&$-0.24362\times10^{-3}\leq \alpha_{s} \leq -0.11501\times 10^{-3}$\\
\hline \\
65&$0.9$&&$47.3\leq n\leq52.5$
&$-0.20854\times10^{-3}\leq \alpha_{s} \leq -0.12598\times 10^{-3}$\\
\hline \hline\\
70& $0.1$&&$45.15\leq
n\leq 55.95$&$-0.34392\times10^{-3}\leq \alpha_{s} \leq -0.11006\times 10^{-3}$\\
\hline \\
70 &$0.4$&&$46.2 \leq n\leq 55.85$
&$-0.29688\times10^{-3}\leq \alpha_{s} \leq -0.11047\times 10^{-3}$\\
\hline \\
70& $0.7$&&$47.8\leq n\leq 55.4$
&$-0.23506\times10^{-3}\leq \alpha_{s} \leq -0.11278\times 10^{-3}$\\
\hline \\
70&$0.9$&&$49.0\leq n\leq 54.2$
&$-0.20467\times10^{-3}\leq \alpha_{s} \leq -0.12484\times 10^{-3}$\\
\hline \hline
\end{tabular}
\end{small}
\end{table*}

\begin{table*}
\begin{small}
\caption{\small{\label{tab:8} The ranges of $n$ in which both
the scalar spectral index and tensor-to-scalar ratio in the power-law
MDBI inflation with constant sound speed are consistent with
68\% CL of the Planck2018 TT, TE, EE+lowE+lensing data. This table
also shows the constraints on the running of the scalar spectral
index obtained from the constraints on $n$.}}
\begin{tabular}{cccccc}
\\ \hline \hline \\$N$& $c_{s}$&&$n$&$\alpha_{s}$\\
\hline \\50& $0.1$&&$39.65\leq
n\leq 45.68$&$-0.31403\times10^{-3}\leq \alpha_{s} \leq -0.15690\times 10^{-3}$\\
\hline \\
50 &$0.4$&&$41.15\leq n\leq 44.8$
&$-0.25848\times10^{-3}\leq \alpha_{s} \leq -0.17134\times 10^{-3}$\\
\hline \\50& $0.7$&&not consistent
&not consistent\\
\hline \\ 50&$0.9$&&not consistent
&not consistent\\
\hline \hline\\
55& $0.1$&&$41.4\leq
n\leq 47.6$&$-0.30811\times10^{-3}\leq \alpha_{s} \leq -0.15303\times 10^{-3}$\\
\hline \\
55 &$0.4$&&$43.0\leq n\leq 46.7$
&$-0.25128\times10^{-3}\leq \alpha_{s} \leq -0.16717\times 10^{-3}$\\
\hline \\55& $0.7$&&not consistent
&not consistent\\
\hline \\ 55&$0.9$&&not consistent
&not consistent\\
\hline \hline\\
60& $0.1$&&$43.18\leq
n\leq 49.45$&$-0.29979\times10^{-3}\leq \alpha_{s} \leq -0.14984\times 10^{-3}$\\
\hline \\
60 &$0.4$&&$44.7\leq n\leq 48.6$
&$-0.24780\times10^{-3}\leq \alpha_{s} \leq -0.16234\times 10^{-3}$\\
\hline \\60& $0.7$&&not consistent
&not consistent\\
\hline \\ 60&$0.9$&&not consistent
&not consistent\\
\hline \hline\\
65& $0.1$&&$44.9\leq
n\leq 51.3$&$-0.29447\times10^{-3}\leq \alpha_{s} \leq -0.14642\times 10^{-3}$\\
\hline \\
65 &$0.4$&&$46.5\leq n\leq 50.4$
&$-0.24100\times10^{-3}\leq \alpha_{s} \leq -0.15921\times 10^{-3}$\\
\hline \\65& $0.7$&&not consistent
&not consistent\\
\hline \\ 65&$0.9$&&not consistent
&not consistent\\
\hline \hline\\
70& $0.1$&&$46.58\leq
n\leq 53.05$&$-0.28944\times10^{-3}\leq \alpha_{s} \leq -0.14437\times 10^{-3}$\\
\hline \\
70 &$0.4$&&$48.2 \leq n\leq 52.1$
&$-0.23676\times10^{-3}\leq \alpha_{s} \leq -0.15722\times 10^{-3}$\\
\hline \\70& $0.7$&&not consistent
&not consistent\\
\hline \\ 70&$0.9$&&not consistent
&not consistent\\
\hline \hline
\end{tabular}
\end{small}
\end{table*}

\section{Confrontation with BICEP2/Keck Array 2014 and Planck2018 data}

Planck2018 data are really powerful in constraining the
cosmological parameters. However, the constraints on the parameters
are somewhat model dependent. For instance, and as we have mentioned
in the Introduction, the constraints on the scalar spectral index
and tensor-to-scalar ratio, $r$, are rather different in the
$\Lambda$CDM$+r$ and $\Lambda$CDM$+r+\frac{dn_{s}}{d\ln k}$
models~\cite{pl18b}. Especially, when the variation of the scalar
spectral index is considered, the upper limit of $r$ is
larger~\cite{pl18b}. To reduce the degeneracies of the
tensor-to-scalar ratio with other cosmological parameters, the
Planck2018 team have used the B-mode polarization data from
BICEP2/Keck Array 2014~\cite{Ade16}. By using the BICEP/Keck
Array-Planck joint cross-correlation, the Planck2018 collaboration
have obtained the tighter constraint on the tensor-to-scalar
ratio~\cite{pl18b}. In this regard, from Planck2018
TT, TE, EE+lowE+lensing+BK14+BAO data, for the $\Lambda$CDM$+r$ model
we have~\cite{pl18b}
\begin{equation}
\label{eq72} r<0.065\,\quad and \quad\, n_{s}=0.9670\pm 0.0037
\end{equation}

and for $\Lambda$CDM$+r+\frac{dn_{s}}{d\ln k}$ we have~\cite{pl18b}
\begin{equation}
\label{eq73} r<0.072\,\quad and \quad\, n_{s}=0.9658 \pm 0.0038
\end{equation}

Our consideration shows that non of the DBI and MDBI models with
varying sound speed are consistent with Planck2018
TT, TE, EE+lowE+lensing+BK14+BAO data, however, these models with
\emph{constant sound speed} are consistent with the mentioned joint data.
We use Planck2018 TT, TE, EE+lowE+lensing+BK14+BAO data at 68$\%$ CL
and 95$\%$ CL to find the the parameter space of $c_{s}$ and $n$
leading to the observationally viable values of $r-n_{s}$ in the
power-law MDBI model. The results are shown in figure 16.

\begin{figure}[]
\begin{center}
\includegraphics[scale=0.29]{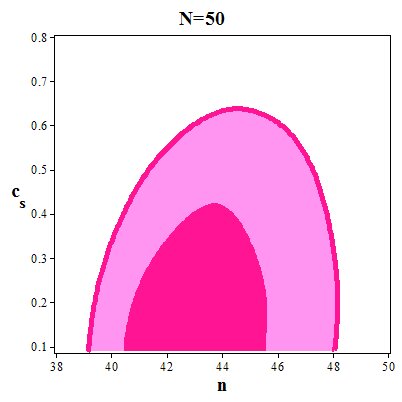}
\includegraphics[scale=0.29]{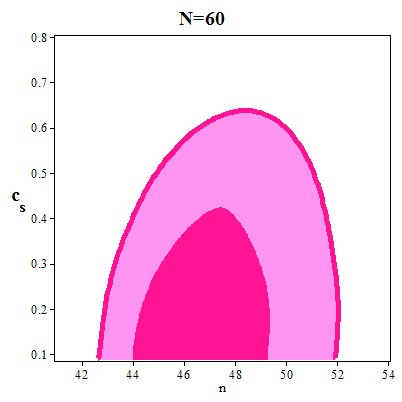}
\includegraphics[scale=0.3]{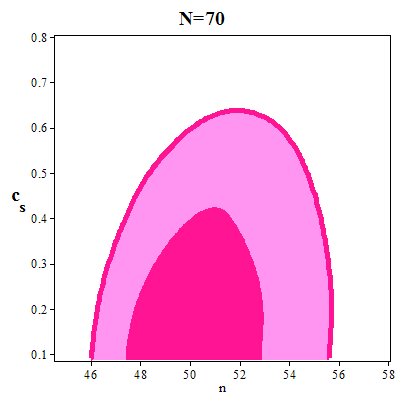}
\end{center}
\caption{\small {Ranges of parameters $c_{s}$ and $n$
in which the scalar spectral index and tensor-to-scalar ratio in a
power-law MDBI inflation with a constant sound speed fulfill
the constraints obtained from Planck2018 TT, TE, EE+lowE+lensing+BK14+BAO at 68\% CL (the dark magenta region) and 95\% CL
(the light magenta region).}}
\label{fig13}
\end{figure}

Here also, we obtain more specific numerical constraints on $n$ and
use them to explore the running of the scalar spectral index. In
this regard, we study $r-n_{s}$ plane of the model in the background
of Planck2018 TT, TE, EE+lowE+lensing+BK14+BAO data. The adopted
values of $c_{s}$ are the same as the ones in previous section.
Figure 17 shows the tensor-to-scalar ratio versus the scalar
spectral index for the power-law DBI (left panel) and power-law MDBI
(right panel) models. By a numerical analysis we have obtained some
constraints on $n$ in which $r-n_{s}$ plane in these two models lies
on the region of 68$\%$ CL and 95$\%$ CL of the Planck2018 TT, TE,
EE+lowE+lensing+BK14+BAO data. These constraints are presented in
tables 9-12.

\begin{figure}[]
\begin{center}
\includegraphics[scale=0.29]{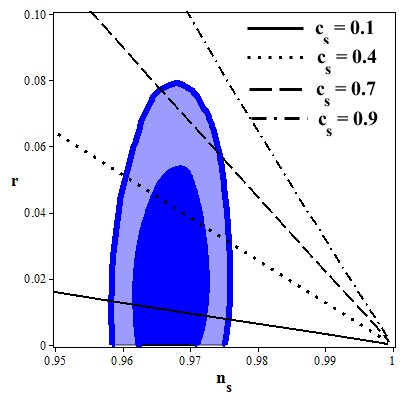}
\includegraphics[scale=0.29]{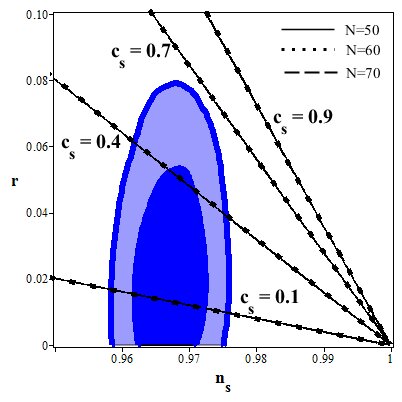}
\end{center}
\caption{\label{fig11}\small {Running of the scalar spectral index
versus the scalar spectral index for the power-law DBI inflation
(left panel) and power-law MDBI inflation (right panel) with
a constant sound speed in the background of the Planck2018
TT, TE, EE+lowE+lensing+BK14+BAO data. }}
\end{figure}

\begin{table*}
\begin{small}
\caption{\small{\label{tab:9} The ranges of $n$ in which both
the scalar spectral index and tensor-to-scalar ratio in the
power-law DBI inflation with a constant sound speed are consistent
with 95\% CL of the Planck2018 TT, TE, EE+lowE+lensing+BK14+BAO data.
This table also shows the constraints on the running of the scalar
spectral index obtained from the constraints on $n$.}}
\begin{tabular}{ccccc}
\\ \hline \hline \\ $c_{s}$&$f_{NL}$&$n$&$\alpha_{s}$\\
\hline \\ $0.1$&$-32.08333$&$119.5\leq
n\leq 205.95$&$-0.52282\times10^{-3}\leq \alpha_{s} \leq -0.17632\times 10^{-3}$\\
\hline \\ $0.4$&$-1.70139$&$126.1\leq n\leq 206$
&$-0.46914\times10^{-3}\leq \alpha_{s} \leq -0.17611\times 10^{-3}$\\
\hline \\ $0.7$&$-0.33730$&$145.1\leq n\leq 195$
&$-0.35280\times10^{-3}\leq \alpha_{s} \leq -0.19580\times 10^{-3}$\\
\hline \\ $0.9$&$-0.07602$&not consistent
&not consistent\\
\hline \hline
\end{tabular}
\end{small}
\end{table*}

\begin{table*}
\begin{small}
\caption{\small{\label{tab:10} The ranges of $n$ in which both
the scalar spectral index and tensor-to-scalar ratio in the
power-law DBI inflation with a constant sound speed are consistent
with 68\% CL of the Planck2018 TT, TE, EE+lowE+lensing+BK14+BAO data.
This table also shows the constraints on the running of the scalar
spectral index obtained from the constraints on $n$.}}
\begin{tabular}{ccccc}
\\ \hline \hline \\ $c_{s}$&$f_{NL}$&$n$&$\alpha_{s}$\\
\hline \\ $0.1$&$-32.08333$&$130.9\leq
n\leq 179.1$&$-0.43586\times10^{-3}\leq \alpha_{s} \leq -0.23306\times 10^{-3}$\\
\hline \\ $0.4$&$-1.70139$&$141.3\leq n\leq 176.8$
&$-0.37381\times10^{-3}\leq \alpha_{s} \leq -0.23897\times 10^{-3}$\\
\hline \\ $0.7$&$-0.33730$&not consistent
&not consistent\\
\hline \\ $0.9$&$-0.07602$&not consistent
&not consistent\\
\hline \hline
\end{tabular}
\end{small}
\end{table*}

\begin{table*}
\begin{small}
\caption{\small{\label{tab:11} The ranges of $n$ in which both
the scalar spectral index and tensor-to-scalar ratio in the power-law
MDBI inflation with constant sound speed are consistent
with 95\% CL of the Planck2018 TT, TE, EE+lowE+lensing+BK14+BAO data.
This table also shows the constraints on the running of the scalar
spectral index obtained from the constraints on $n$.}}
\begin{tabular}{cccccc}
\\ \hline \hline \\$N$& $c_{s}$&&$n$&$\alpha_{s}$\\
\hline \\50& $0.1$&&$39.16\leq
n\leq 48.08$&$-0.33491\times10^{-3}\leq \alpha_{s} \leq -0.12342\times 10^{-3}$\\
\hline \\
50 &$0.4$&&$40.4\leq n\leq 47.8$
&$-0.28310\times10^{-3}\leq \alpha_{s} \leq -0.12646\times 10^{-3}$\\
\hline \\
50& $0.7$&&not consistent
&not consistent\\
\hline \\
50&$0.9$&&not consistent
&not consistent\\
\hline \hline\\
55& $0.1$&&$40.90\leq
n\leq 50.0$&$-0.32817\times10^{-3}\leq \alpha_{s} \leq -0.12076\times 10^{-3}$\\
\hline \\
55 &$0.4$&&$42.19\leq n\leq 49.74$
&$0.27690\times10^{-3}\leq \alpha_{s} \leq -0.12335\times 10^{-3}$\\
\hline \\
55& $0.7$&&not consistent
&not consistent\\
\hline \\
55&$0.9$&&not consistent
&not consistent\\
\hline \hline\\
60& $0.1$&&$42.7\leq
n\leq 51.93$&$-0.31891\times10^{-3}\leq \alpha_{s} \leq -0.11777\times 10^{-3}$\\
\hline \\
60 &$0.4$&&$43.95\leq n\leq 51.6$
&$-0.27040\times10^{-3}\leq \alpha_{s} \leq -0.12097\times 10^{-3}$\\
\hline \\
60& $0.7$&&not consistent
&not consistent\\
\hline \\
60&$0.9$&&not consistent
&not consistent\\
\hline \hline\\
65& $0.1$&&$44.33\leq
n\leq53.81$&$-0.31548\times10^{-3}\leq \alpha_{s} \leq -0.11532\times 10^{-3}$\\
\hline \\
65 &$0.4$&&$45.68\leq n\leq 53.51$
&$-0.26469\times10^{-3}\leq \alpha_{s} \leq -0.11790\times 10^{-3}$\\
\hline \\
65& $0.7$&&not consistent
&not consistent\\
\hline \\
65&$0.9$&&not consistent
&not consistent\\
\hline \hline\\
70& $0.1$&&$46.02\leq
n\leq 55.60$&$-0.30923\times10^{-3}\leq \alpha_{s} \leq -0.11373\times 10^{-3}$\\
\hline \\
70 &$0.4$&&$47.37 \leq n\leq 55.31$
&$-0.25966\times10^{-3}\leq \alpha_{s} \leq -0.11600\times 10^{-3}$\\
\hline \\
70& $0.7$&&not consistent
&not consistent\\
\hline \\
70&$0.9$&&not consistent
&not consistent\\
\hline \hline
\end{tabular}
\end{small}
\end{table*}

\begin{table*}
\begin{small}
\caption{\small{\label{tab:12} The ranges of $n$ in which both
scalar spectral index and tensor-to-scalar ratio in the power-law
MDBI inflation with a constant sound speed are consistent
with 68\% CL of the Planck2018 TT, TE, EE+lowE+lensing+BK14+BAO data.
This table also shows the constraints on the running of the scalar
spectral index obtained from the constraints on $n$.}}
\begin{tabular}{cccccc}
\\ \hline \hline \\$N$& $c_{s}$&&$n$&$\alpha_{s}$\\
\hline \\50& $0.1$&&$40.42\leq
n\leq 45.57$&$-0.28504\times10^{-3}\leq \alpha_{s} \leq -0.15878\times 10^{-3}$\\
\hline \\
50 &$0.4$&&$42.7\leq n\leq 44.51$
&$-0.21590\times10^{-3}\leq \alpha_{s} \leq -0.17677\times 10^{-3}$\\
\hline \\50& $0.7$&&not consistent
&not consistent\\
\hline \\ 50&$0.9$&&not consistent
&not consistent\\
\hline \hline\\
55& $0.1$&&$42.24\leq
n\leq 47.46$&$-0.27768\times10^{-3}\leq \alpha_{s} \leq -0.15516\times 10^{-3}$\\
\hline \\
55 &$0.4$&&$44.58\leq n\leq 46.37$
&$-0.20992\times10^{-3}\leq \alpha_{s} \leq -0.17298\times 10^{-3}$\\
\hline \\55& $0.7$&&not consistent
&not consistent\\
\hline \\ 55&$0.9$&&not consistent
&not consistent\\
\hline \hline\\
60& $0.1$&&$43.98\leq
n\leq 49.29$&$-0.27264\times10^{-3}\leq \alpha_{s} \leq -0.15220\times 10^{-3}$\\
\hline \\
60 &$0.4$&&$46.4\leq n\leq 48.2$
&$-0.20487\times10^{-3}\leq \alpha_{s} \leq -0.16918\times 10^{-3}$\\
\hline \\60& $0.7$&&not consistent
&not consistent\\
\hline \\ 60&$0.9$&&not consistent
&not consistent\\
\hline \hline\\
65& $0.1$&&$45.72\leq
n\leq 51.12$&$-0.26718\times10^{-3}\leq \alpha_{s} \leq -0.14898\times 10^{-3}$\\
\hline \\
65 &$0.4$&&$48.15\leq n\leq 49.98$
&$-0.20101\times10^{-3}\leq \alpha_{s} \leq -0.16596\times 10^{-3}$\\
\hline \\65& $0.7$&&not consistent
&not consistent\\
\hline \\ 65&$0.9$&&not consistent
&not consistent\\
\hline \hline\\
70& $0.1$&&$47.39\leq
n\leq 52.88$&$-0.26329\times10^{-3}\leq \alpha_{s} \leq -0.14667\times 10^{-3}$\\
\hline \\
70 &$0.4$&&$49.84 \leq n\leq 51.79$
&$-0.20052\times10^{-3}\leq \alpha_{s} \leq -0.16368\times 10^{-3}$\\
\hline \\70& $0.7$&&not consistent
&not consistent\\
\hline \\ 70&$0.9$&&not consistent
&not consistent\\
\hline \hline
\end{tabular}
\end{small}
\end{table*}

\newpage

\section{Summary and Conclusion}

In the mimetic gravity, to have a non-zero sound speed, one should
consider higher order terms as $\gamma\square \phi$ in the action. However, this
model suffers from gradient instabilities. By considering the direct
couplings of the higher derivatives of the mimetic field to the
curvature of the space-time, one can overcome the instabilities
issue in some regions of the models' parameters space. In this paper, instead
of including the higher order derivatives of the scalar field, we
have proposed a new model by adding a DBI like term as $f^{-1}\sqrt{1-f\dot{\phi}^{2}}$ in
the action of the simple mimetic model with a potential. By adding
this term, we can have an instabilities-free mimetic model, at least
in some regions of the model's parameters space. Indeed, in this paper, we
have considered both the DBI and our newly proposed Mimetic DBI (MDBI)
models in details and for the sake of comparison. We have
studied the power-law inflation in these models and compared the
results with Planck2018 data to seek for the observational viability
of the models. In this regard, we have considered two general cases:
varying sound speed and constant sound speed. For the case of varying sound
speed, by studying $r-n_{s}$ and $\alpha-n_{s}$ planes and comparing
with Planck2018 observational data, we have found some constraints
on the parameter $n$ where $a=a_{0}\,t^{n}$. The constraints on $n$ gave us the observationally viable
values of $c_{s}$ and $f_{NL}$. For the constant sound speed, we
have used the observationally viable values of $f_{NL}$ to find the
viable values of the sound speed. Then, with the viable values of
the sound speed, we have found some constraints on $n$ leading to
the observationally viable values of $r$, $n_{s}$ and $\alpha_{s}$.

In the varying sound speed case, we have firstly studied the DBI
model. We have obtained the main background equations in this setup.
By using these equations, we have obtained the perturbation
parameters such as the scalar spectral index, its running and
tensor-to-scalar ratio in this setup and also the corresponding
potential in terms of the Hubble parameters and its derivatives. By
studying the power-law inflation in this model and constructing the
corresponding potential, we have shown that this model is not
consistent with Planck2018 TT, TE, EE+lowE+lensing data. Note that,
although $r-n_{s}$ plane in this model with $V_{-}$ is consistent
with Planck2018 TT, TE, EE+lowE+lensing data at the 68$\%$ CL and 95$\%$
CL, but the constraints on $n$ lead to the values of $f_{NL}$ which
are too large to be consistent with observational data.
Then we have proceeded to see the situation in the mimetic DBI scenario.

In the MDBI model, we firstly obtained the main background equations of model.
We have also found the Lagrange multiplier and potential in this setup in terms of the Hubble
parameter. By numerical studying of the power-law MDBI model, we
have shown that this model, in some regions of the model's parameters space,
is free of the ghost and gradient instabilities. By obtaining the
the perturbation's parameters in this model and comparing the
results with the observational data, we have found that our MDBI
model is consistent with Planck2018 TT, TE, EE+lowE+lensing data at the
95$\%$ CL. In this case, the constraints on $n$ lead to the small
amplitude of the non-Gaussianity which is consistent with
observation.

In the constant sound speed case also, we have studied both DBI and MDBI
inflation with power-law scale factor. We have obtained the
potential, the DBI function ($f$) and the slow-roll parameters in terms of
$c_{s}$, $N$ and $n$. We have adopted some sample values of the
the sound speed allowed by the Planck2015 data. By these adopted values of
the sound speed we have studied $r-n_{s}$ and $\alpha_{s}-n_{s}$ planes
numerically and have obtained some constraints on parameter $n$. Both the DBI and
MDBI models with power-law scale factor are consistent with
Planck2018 TT, TE, EE+lowE+lensing data at the 68$\%$ CL and 95$\%$ CL.
However, for the DBI model, the constraint on $n$ is as $n\sim
10^{2}$ while, for the newly proposed MDBI model it is as $n\sim 10^{1}$. We have
also obtained the parameters space of $c_{s}$ and $n$ in which
the scalar spectral index and tensor-to-scalar ratio in a power-law
MDBI inflation with constant sound speed fulfill the constraints
obtained from Planck2018 TT, TE, EE+lowE+lensing at the 68\% CL and 95\%
CL. According to our analysis, the constraint on the constant sound
speed in the power-law MDBI inflation is as $c_{s}\leq 0.95$.

\begin{table*}
\begin{small}
\caption{\small{\label{tab:13} Observational viability of the
models considered in this paper.}}
\begin{tabular}{ccccc}
\\ \hline \hline \\ Model &Sound Speed &Planck2018 TT, TE, EE+lowE&Planck2018 TT, TE, EE+lowE\\
&&+lensing&+lensing+BK14+BAO\\
\hline
\\ DBI&varying&not consistent&not consistent\\
\hline \\ MDBI&varying&consistent (with $n\sim 10^{1}$)
&not consistent\\
\hline \\ DBI&constant&consistent (with $n\sim 10^{2}$)
&consistent (with $n\sim 10^{2}$)
&\\
\hline \\ MDBI&constant&consistent (with $n\sim 10^{1}$)
&consistent (with $n\sim 10^{1}$)\\
\hline \hline
\end{tabular}
\end{small}
\end{table*}

Finally, we note that to reduce the degeneracies of the
tensor-to-scalar ratio with other cosmological parameters and obtain
the tighter constraint on the tensor-to-scalar ratio, Planck2018 collaboration
has used the B-mode polarization data from BICEP2/Keck Array 2014.
In this regard, we have explored the DBI and newly proposed MDBI model in
confrontation with Planck2018 TT, TE, EE+lowE +lensing+BK14+BAO data,
where BK14 refers to BICEP2/Keck Array 2014 data. The results show that
in this case also, the constraint on $n$ for the DBI model is as
$n\sim 10^{2}$ and for the MDBI model it is as $n\sim 10^{1}$.
However, by obtaining the parameters space of $c_{s}$ and $n$ in
which the scalar spectral index and tensor-to-scalar ratio in a
power-law MDBI inflation are consistent with Planck2018
TT, TE, EE+lowE+lensing+BK14+BAO at 68\% CL and 95\% CL, we have found
a tighter constraint on the constant sound speed as $c_{s}\leq
0.64$. Table 13 summarizes the final results of this paper.\\

\textbf{Appendix: The Third Slow-Roll Parameter in the Mimetic DBI (MDBI)
Model}

\begin{eqnarray}
s=\frac{{\sqrt {2}\,\kappa}^{5}}{4}\Bigg[\frac {  54\,
H^{5}(N)H''(N)+162 {H'}^{2}(N)H^{4}(N)+144H^{4}(N )H'(N) H''(N)}{
\Big( -H'(N)H(N) \Big) ^{3/2}  \Big( 3\,  H ^{2} ( N ) +2\,H \left(
N \right) H'(N)-{\kappa}^{2} \Big) J}\nonumber\\
+\frac {288 {H'}^{3}(N) H^{3} ( N )   +120H^{3} ( N ){
H'}^{2}(N)H''(N)-27H^{3} ( N ) H''(N){\kappa}^{2}+168{H'}^{4}(N)
H^{2}( N )}{ \Big( -H'(N)H(N) \Big) ^{3/2}  \Big( 3\,  H ^{2} ( N )
+2\,H \left(
N \right) H'(N)-{\kappa}^{2} \Big) J}\nonumber\\
+\frac {32 H^{2} ( N){H'}^{3}(N)H''(N)+9 H^{2}( N )
{H'}^{2}(N){\kappa}^{2} -24H^{2} ( N ) H'(N)H''(N)\,{\kappa}^{2}}{
\Big( -H'(N)H(N) \Big) ^{3/2}  \Big( 3\,  H ^{2} ( N ) +2\,H \left(
N \right) H'(N)-{\kappa}^{2} \Big) J}\nonumber\\
+\frac {32\,{H'}^{5}(N) H( N) -4H ( N ) {H'}^{2}(N)H''(N)\,
{\kappa}^{2}+3H ( N )
H''(N)\,{\kappa}^{4}-4{H'}^{4}(N){\kappa}^{2}-3{H'}^{2}(N){\kappa}^{4}
}{ \Big( -H'(N)H(N) \Big) ^{3/2}  \Big( 3\,  H ^{2} ( N ) +2\,H
\left( N \right) H'(N)-{\kappa}^{2} \Big) J}\Bigg]\nonumber
\end{eqnarray}

with

\begin{eqnarray}
J=54H'(N)\,  H^{5} ( N ) +108\,{H'}^{2}(N) H^{4}(N) +72{H'}^{3}(N)
H^{3} ( N )  -54 H^{3} ( N
)H'(N)\,{\kappa}^{2}\nonumber\\+16{H'}^{4}(N) H^{2} ( N ) -72 H^{2}
( N ) {H'}^{2}(N){\kappa}^{2}-18  H^{2} ( N ) {\kappa}^{4}-24H ( N )
{H'}^{3}(N){\kappa}^{2}\nonumber\\-6H( N)
H'(N)\,{\kappa}^{4}+4\,{H'}^{2}(N){\kappa}^{4}+3\,{\kappa}^{6}\nonumber
\end{eqnarray}

{\bf Acknowledgement}\\
This work has been supported financially by Research Institute for
Astronomy and Astrophysics of Maragha (RIAAM) under research project
number 1/6275-6.\\

\end{document}